\documentclass[12pt]{article}
\usepackage[cp866]{inputenc}
\usepackage[russian]{babel}
\usepackage{graphicx}

\begin{document}
\begin{center}
{\large\bf On the Various Aspects of
Hamiltonian Description of the  Mechanics of Continuous Media} \\
\vskip 1cm
{ G.P.Pronko}\\

{\it Institute for
High Energy Physics , Protvino, Moscow reg., Russia}
\end{center}

\begin{abstract}
We consider a general approach to the theory of continuous media starting from Lagrangian formalism. This formalism which uses the trajectories if constituents of media is very convenient for taking into account different types of interaction between particles typical for different media. Building the Hamiltonian formalism we discuss some issues which is not very well known, such as relation of famous Thompson theorem with the symmetry with respect to volume preserving diffeomorphisms. We also discuss the relation between  Euler and Lagrange description and present similar to Euler $C^2$ formulation of continuous mechanics. In these general frameworks we consider as examples the theory of plasma and gravitating gas.
\end{abstract}

\section{Introduction}

At the present time almost all fundamental physical phenomena could be formulated  in the frameworks of either classical or quantum mechanics. That means that these phenomena admits the Hamiltonian description, which due to its long history developed many powerful methods of analysis of the
general properties of evolution of the systems and the tools for the solutions of partial problems.

In this respect the fluid mechanics stands aside (in spite of its name) from orthodox mechanics. The reasons for that is not only the infinite number of degrees of freedom of fluid which could be treated e.g. within classical field theory or statistical mechanics. The main difference between conventional field theory and continuous mechanics is that in the first case we can speak about the dynamics of the field at one point in space (which of course interacts with the field at the neighboring points), while in the
case of the fluid, describing the interaction of the neighboring
particles which constitute the fluid  we are loosing its position
in the space due to the motion of fluid. In the same time the
objective of usual problems of hydrodynamics is to define the
velocity, density and a thermodynamical variable (pressure or
entropy) as the functions of the coordinates $\vec x$ and time $t$ for the appropriate boundary conditions and/or initial data \cite{Landau}. The similar problems also appear for
magnetohydrodynamics dealing with sufficiently dense plasma
\cite{Chen}. For the developing of the Hamiltonian formalism we need to start with more detailed description based initially on the trajectories of constituents of media. This description is especially important for plasma,
because the fundamental electromagnetic interaction could be
formulated only in terms of the trajectories of the charges.
Needless to say that some aspects of this approach was extensively
studied in the series of papers by J.Marsden, A.Weinstein,
P.Kupershmidt, D.Nolm , T.Ratiu and C.Levermore \cite{Mars}
especially in the context of stability problem.

Of course not all properties of fluid could be formulated in the frameworks of the Hamiltonian approach. For example, we leave open the question of the energy dissipation, viscosity et cetera.

\section{The Lagrangian equations of motion}

In fluid mechanics there are two different pictures of
description. The first, usually refereed as Eulerian, uses as the
coordinates  the space dependent fields of velocity, density and
some thermodynamic variable. The second, Lagrangian description,
uses the coordinates of the particles $\vec x(\xi_i,t)$  labeled
by the set of the parameters $\xi_i$, which could be considered as
the initial positions $\vec\xi=\vec x(t=0)$
 and time $t$. These initial positions  $\vec \xi$ as well, as the
coordinates $\vec x(\xi_i,t)$ belong to some domain $D \subseteq
R^3$.
In sequel we shall consider only conservative systems, where the
paths of
different particles do not cross, therefore it is clear  that  the
functions $\vec x(\xi_i,t)$ define a diffeomorphism of  $D \subseteq
R^3$
and the inverse functions $\vec \xi(x_i,t)$ should also exist.
\begin{eqnarray}\label{1}
x_j(\xi_i,t)\Big|_{\vec \xi=\vec\xi(x_i,t)}&=x_j,\nonumber\\
\xi_j(x_i,t)\Big|_{\vec x=\vec x(\xi_i,t)}&=\xi_j.
\end{eqnarray}
The density of the particles in space at time $t$ is
\begin{equation}\label{2}
\rho(\vec x,t)=\int d^3 \xi \rho_0(\xi_i)\delta(\vec x-\vec
x(\xi_i,t)),
\end{equation}
where $\rho_0(\xi)$ is the initial density at time $t=0$.
The velocity field  $\vec v$ as a function of coordinates $\vec x$
and $t$
is:
\begin{equation}\label{3}
\vec v(x_i,t)=\dot{\vec x}(\vec\xi(x_i,t),t),
\end{equation}
where $\vec\xi(x,t)$ is the inverse function (\ref{1}). The velocity
also
could be written in the following form:
\begin{equation}\label{4}
\vec v(x_i,t)=\frac{\int d^3 \xi \rho_0(\xi_i)\dot{\vec
x}(\xi_i,t)\delta(\vec
x-\vec x(\xi_i,t))}{\int d^3 \xi \rho_0(\xi_i)\delta(\vec x-\vec
x(\xi_i,t))},
\end{equation}
or
\begin{equation}\label{5}
\rho (x_i,t)\vec v(x_i,t)=\int d^3 \xi \rho_0(\xi_i)\dot{\vec
x}(\xi_i,t)\delta(\vec x-\vec x(\xi_i,t)).
\end{equation}
Let us calculate the time derivative of the density using its
definition
(\ref{2}) :
\begin{eqnarray}\label{6}
&\dot \rho (x_i,t)=\displaystyle\int d^3 \xi
\rho_0(\xi_i)\frac{\partial}{\partial t}
\delta(\vec x-\vec x(\xi_i,t))\nonumber\\
&=\displaystyle\int d^3 \xi \rho_0(\xi_i)\left(-\dot{\vec
x}(\xi_i,t)\right)\frac{\partial}{\partial \vec x}
\delta(\vec x-\vec x(\xi_i,t))\nonumber\\
&=-\frac{\partial}{\partial \vec x}\int d^3 \xi
\rho_0(\xi_i)\dot{\vec x}(\xi_i,t)\delta(\vec x-\vec
x(\xi_i,t))\nonumber\\
&=-\displaystyle\frac{\partial}{\partial \vec x}\rho (x_i,t)
\vec v (x_i,t)
\end{eqnarray}
In such a way we verify the continuity equation of fluid dynamics:
\begin{equation}\label{7}
\dot \rho (x_i,t)+\vec\partial\Bigl(\rho (x_i,t)\vec v
(x_i,t)\Bigr)=0.
\end{equation}.

Using the coordinates $\vec x(\xi_i,t)$ as a configurational
variables we
can consider the simplest motion of the fluid described by the
Lagrangian
\begin{equation}\label{8}
L=\int d^3 \xi \rho_0(\xi_i)\frac{m\dot{\vec x}^2 (\xi_i,t)}{2}.
\end{equation}
The equations of motion which follow from (\ref{8}) apparently are
\begin{equation}\label{9}
m\rho_0(\xi_i)\ddot{\vec x} (\xi_i,t)=0
\end{equation}
Now let us find what does this equation mean for the density and
velocity
of the fluid. For that we shall differentiate both sides of (\ref{5})
with
respect to time
\begin{eqnarray}\label{10}
&\displaystyle\frac{\partial}{\partial t}\rho (x_i,t)\vec
v(x_i,t)=\int
d^3 \xi \rho_0(\xi_i)\ddot{\vec x}(\xi_i,t)\delta(\vec x-\vec
x(\xi_i,t))\nonumber\\
&+\int d^3 \xi
\rho_0(\xi_i)\dot{\vec
x}(\xi_i,t)\displaystyle\frac{\partial}{\partial
t}\delta(\vec x-\vec x(\xi_i,t))
\end{eqnarray}
The first term in the r.h.s. of (\ref{10}) vanishes due to the
equations
of motion(\ref{9}) and transforming the second in the same way, as we
did
in (\ref{6}) we arrive at
\begin{equation}\label{11}
\displaystyle\frac{\partial}{\partial t} \rho (x_i,t)\vec v(x_i,t)+
\displaystyle\frac{\partial}{\partial x_k}\Bigl(\rho (x_i,t)
\vec v (x_i,t) v_k(x_i,t)\Bigr)=0
\end{equation}
Let us rewrite (\ref{11}) in the following form:
\begin{eqnarray}\label{12}
&\vec v(x_i,t)\Big[\dot \rho
(x_i,t)+\displaystyle\frac{\partial}{\partial
x_k}\Bigl(\rho (x_i,t) v_k (x_i,t)\Bigr)\Big]\nonumber\\
&+\rho(x_i,t)\Bigl[\dot{\vec
v}(x_i,t)+v_k(x_i,t)\displaystyle\frac{\partial}{\partial
x_k} \vec v (x_i,t)\Bigr]=0.
\end{eqnarray}
The first term in (\ref{12}) vanishes due to the continuity equation,
while the second gives Euler's equation in the case of the free flow:
\begin{equation}\label{13}
\dot{\vec v}(x_i,t)+
v_k(x_i,t)\displaystyle\frac{\partial}{\partial
x_k} \vec v (x_i,t)=0
\end{equation}

Before moving further we need to make one comment concerning the
Lagrangian (\ref{8}). The reader may get an impression that this
Lagrangian depends not only on the dynamical variables at the time
$t$, but also at the initial time $t=0$ through the presence of
$\rho_0(\xi_i)$. In order to eliminate the doubt let us take the
unity
\begin{equation}\label{a}
1=\int d^3 x \delta(\vec x-\vec x(\xi_i,t))
\end{equation}
and insert it into the integrand of (\ref{8}).
\begin{equation}\label{b}
L=\int d^3 \xi d^3 x\rho_0(\xi_i)\frac{m\dot{\vec x}^2
(\xi_i,t)}{2}\delta(\vec x-\vec x(\xi_i,t)).
\end{equation}
Performing the integration in (\ref{b}) over $\xi$ we obtain:
\begin{equation}\label{c}
L=\int d^3 x\rho(x_i,t)\frac{m\vec v(x_i,t)^2 }{2}.
\end{equation}
This representation of the Lagrangian (\ref{8}) shows explicitly the
dependence of $L$ only of the dynamical variables at the time $t$
(besides we are not going  to consider here the Euler variables as an
independent).

In order to have more realistic model of the fluid
we need to introduce into the Lagrangian (\ref{8})
the "potential energy" term which will give rise to the  internal
pressure field in Euler's equation.

As we have mentioned above,
the functions $\vec x(\xi_i,t)$ define a diffeomorphism in $R^3$
therefore the  matrix
\begin{equation}\label{14}
A^j_k(\xi_i,t)=\frac{\partial x_j(\xi_i,t)}{\partial \xi_k}
\end{equation}
is non-degenerate for any $\xi_i$ and $t$. The integral (\ref{1}) may
be expressed via the Jacobean -- the determinant of $A^j_k(\xi_i,t)$:
\begin{equation}\label{15}
\rho(\vec x,t)=\frac{\rho_0(\xi_i)}{detA(\xi_i,t)}\Bigg|_{\vec
\xi=\vec \xi(x_i,t)}.
\end{equation}
For some problems one can choose  the initial density $\rho_0(\xi_i)$
to be uniform in $D\subseteq R^3$, and  effectively normalize the
density field $\rho(\xi_i,t)$ by putting $\rho_0(\xi_i)=1$ (one
particle in the elementary volume). However, we prefer to keep
$\rho_0(\xi_i)$ arbitrary, because in the case e.g. the presence of
soliton in the fluid the density could not be uniform at any time.

Taking the "potential energy" to be the  functional of $detA$ we can write
the lagrangian in the following form:
\begin{equation}\label{17}
L=\int d^3 \xi \rho_0(\xi_i)\Biggl[\frac{m\dot{\vec x}^2
(\xi_i,t)}{2}-f(detA(\xi_i,t))\Biggr].
\end{equation}
In order the Lagrangian be the functional of the dynamical variables
at the time $t$, the function $f(detA(\xi_i,t))$, should has the
following form:
\begin{equation}\label{d}
f(detA(\xi_i,t))=V(\frac{\rho_0(\xi_i)}{detA(\xi_i,t)})\frac{detA(\xi
_i,t)}{\rho_0(\xi_i)}
\end{equation}
Now the equations of motion become
\begin{equation}\label{18}
m \rho_0(\xi_i)\ddot{x_j}
(\xi_i,t)-\displaystyle\frac{\partial}{\partial
\xi_k}\Biggl(\rho_0(\xi_i)\left(A^{-1}\right)_j\,^k(\xi_i,t)
f'(detA)detA\Biggr)=0
\end{equation}
Substituting $\ddot{x_j} (\xi_i,t)$ from (\ref{18}) to the
equation(\ref{10}) and acting as we did in the derivation of the
equation
(\ref{13}), we obtain
\begin{equation}\label{19}
m\rho (x_i,t)\displaystyle\left(\frac{\partial}{\partial
t}+v_k(x_i,t) \frac{\partial}{\partial
x_k}\right)v_j(x_i,t)-\displaystyle\frac{\partial}{\partial
x_j}\left(\rho_0(\xi_i)f'(detA(\xi_i,t))\Bigg|_{\vec \xi=\vec
\xi(x_i,t)}\right)=0
\end{equation}

It is now obvious that if we identify the
$-\frac{1}{m} \rho_0(\xi_i)f'(detA(\xi_i,t))\Big|_{\vec \xi=\vec
\xi(x_i,t)}$ with
pressure $p(x_i,t)$, the equation (\ref{19}) takes the form of
usual Euler equation without viscosity:
\begin{equation}\label{20}
\rho (x_i,t)\displaystyle\left(\frac{\partial}{\partial t}+v_k(x_i,t)
\frac{\partial}{\partial
x_k}\right)v_j(x_i,t)=-\displaystyle\frac{\partial}{\partial
x_j}p(x_i,t).
\end{equation}
For the function $f(detA(\xi_i,t))$ parameterized as in (\ref{d}), the
pressure has the following form:
\begin{equation}\label{e}
p(x_i,t)=\frac{\rho(x_i,t)^2}{m}\frac{\partial}{\partial
\rho}(\frac{V(\rho(x_i,t))}{\rho(x_i,t)})
\end{equation}
The pressure $p(x_i,t)$ which appeared here is the result of
interaction between particles, which constitute the fluid. We can
also add  to the Lagrangian (\ref{17}) the term, which describes the
interaction with an external field:
\begin{equation}
L_{ext}=-\int d^3 \xi \rho_0(\xi_i)U(x_i(\xi_i,t)),
\end{equation}
and the Euler equations take the form:
\begin{equation}
\rho (x_i,t)\displaystyle\left(\frac{\partial}{\partial t}+v_k(x_i,t)
\frac{\partial}{\partial
x_k}\right)v_j(x_i,t)=-\displaystyle\frac{\partial p(x_i,t)}{\partial
x_j}-\frac{\rho(x_i,t)}{m}\displaystyle\frac{\partial U(x_i)}
{\partial x_j}.
\end{equation}

As we shall see later  the description in terms of Lagrange variables is
very convenient for introducing interaction between particles because in
this picture the usual coordinates are the dynamical variables. For
example, considering two components fluid -- ions and electron, interacting
with electromagnetic field we obtain the theory of plasma. Introduction the
interaction of particles, which constitute the fluid with color gluon field
gives us the theory of quark-gluon plasma. Very interesting system is the
gravitating gas, where particles interact with each other through
gravitation field. These different types of interaction changes the
physical properties if the media, but all of them could be described within
the common approach.

\section{Hamiltonian formalism}

The description of fluid in terms of Lagrangian variables brings no
difficulties in construction of canonical formalism. Indeed, for the
Lagrangian (\ref{17}), the canonical coordinates will be the
functions $\vec x(\xi_i,t)$ and its conjugated momenta are defined as
the derivatives of the Lagrangian with respect to the velocities
$\dot {\vec x}(\xi_i,t)$:
\begin{equation}\label{1.1}
\vec p(\xi_i,t)=\frac{\delta L}{\delta \dot {\vec x}(\xi_i,t)}=
m\rho_0(\xi_i)\dot
{\vec x}(\xi_i,t),
\end{equation}
The Hamiltonian is given by the Legendre transformation of
the Lagrangian:
\begin{equation}\label{1.2}
H=\int d^3 \xi \left(\frac{1}{2m \rho_0(\xi_i)}\vec
p^2(\xi_i,t)+f(detA(\xi_i,t))\right).
\end{equation}
The canonical Poisson brackets is defined by
\begin{equation}\label{1.3}
\{p_j(\xi_i), x_k(\xi'_i)\}=\delta_{jk}\delta^3(\xi_i-\xi'_i)
\end{equation}
Apparently the Poisson brackets (\ref{1.3}) and the Hamiltonian
(\ref{1.2}) define the  equations of motion for the canonical
variables, which are equivalent to the Lagrange equations. The phase
space $\Gamma$ of the fluid is formed by $\vec x(\xi_i,t),\vec
p(\xi_i,t)$. What is missing at the moment is the $x$-space
interpretation of the variables $\vec x(\xi_i,)$ and $\vec p(\xi_i)$
and the goal of this section is to find a canonical transformation of
$\vec x(\xi_i,)$ and $\vec p(\xi_i)$ to equivalent set of coordinates
in $\Gamma$, which are $x$-dependent functions.

Let us introduce the new objects using the same averaging, as we used
in the previous section
\begin{equation}\label{1.4}
\vec l(x)=\int d^3 \xi\vec p(\xi_i)\delta(\vec x-\vec x(\xi_i))=
\rho(x)\vec p(\xi_i(x)).
\end{equation}
The Poisson brackets of $\vec l(x)$, induced by (\ref{1.3}) has the
following form:
\begin{equation}\label{1.5}
\{l_j(x_i),l_k(y_i)\}=\Bigl[\l_k(x_i)\frac{\partial}{\partial
x_j}+\l_j(y_i)\frac{\partial}{\partial x_k}\Bigr]\delta(\vec x-\vec
y).
\end{equation}
These Poisson brackets was introduced geometrically as "the
hydrodynamic-type brackets" long ago in the papers  \cite
{DubrovinNovikov} without physical explanation. The present
discussion reveals the origin of these brackets. The commutation
relation (\ref{1.5}) coincides with algebra of 3-dimensional
diffeomorphisms. In other words, $\vec l(x)$ are the generators of
the finite  diffeomorphism $x_j\rightarrow \phi_j(x_i)$ of
any $x$-dependent dynamical variable in $\Gamma$. It should be
mentioned that the group of diffeomorphisms, generated by  $l_j(x_i)$
is not a gauge symmetry in case of fluid mechanics, as it is in the
case of e.g. relativistic string or membrane. In the same time in
fluid mechanics there is an infinite dimensional symmetry (but not a
gauge symmetry) with respect to special (i.e. volume preserving)
diffeomorphisms $SDiff$, which will be considered in later.

Let us leave for a moment $\vec l(x)$ and consider another
$x$-dependent functions $\xi_i(x)$ which are the inverse to
$x_j(\xi_i)$ functions (\ref{1}). Differentiating the first equation
(\ref{1}) with respect to $x$ we obtain:
\begin{equation}\label{1.6}
A^j_k(\xi_i(x_k))\frac{\partial \xi_k(x_i)}{\partial x_l}=\delta
^j_l,
\end{equation}
in other words, the matrix
\begin{equation}\label{1.7}
a^k_l(x_i)=\frac{\partial \xi_k(x_i)}{\partial x_l}
\end{equation}
is inverse to $A^j_k(\xi_i(x_k))$. Therefore, from (\ref{15}) it
follows that
\begin{equation}\label{1.8}
\rho(x_i)=\rho_0 (\xi_i(x_i)) det a^k_l(x_i)
\end{equation}
To simplify the calculation of the Poisson brackets we can express
$\xi_j(x_i)$ in the following form:
\begin{equation}\label{1.9}
\xi_j(x_i)=\frac{\int d^3 \xi \rho_0(\xi_i)\xi_j\delta(\vec x-\vec
x(\xi_i,t))}{\int
d^3\xi \rho_0(\xi_i) \delta(\vec x-\vec x(\xi_i,t))}.
\end{equation}
From (\ref{1.9}) we easily obtain
\begin{equation}\label{1.10}
\{\xi_j(x_i),\xi_k(y_i)\}=0
\end{equation}
The calculation of the Poisson brackets between $\xi_j(x_i)$ and
$l_j(x_i)$
is more involved, but the result is simple:
\begin{equation}\label{1.11}
\{l_j(x_i),\xi_k(y_i)\}=-\frac{\partial \xi_k(x_i)}{\partial
x_j}\delta(\vec x-\vec y)
\end{equation}
In such a way we have constructed the set of $x$-dependent
coordinates
$\left(l_j(x_i), \xi_j(x_i)\right)$ in the phase space $\Gamma$, but
the
transformation
\begin{equation}\label{1.12}
\left(p_j(\xi_i),x_j(\xi_i)\right) \rightarrow \left( l_j(x_i),
\xi_j(x_i)\right)
\end{equation}
is not  canonical.  The set of canonical $x$-dependent coordinates in
$\Gamma$ could be obtained in the following way. Let us multiply both
sides of (\ref{1.11}) by matrix $A^j_m(\xi_i(x_k))$:
\begin{equation}\label{1.13}
A^j_m(\xi_i(x_k))\{l_j(x_i),\xi_k(y_i)\}=
-\delta ^k_m\delta(\vec x-\vec y),
\end{equation}
where we have used (\ref{1.6}). Further, due to  the relation
(\ref{1.10}), we can put $A^j_m(\xi_i(x_k))$ inside the brackets and
obtain:
\begin{equation}\label{1.14}
\{\pi_m(x_i),\xi_k(y_i)\}=\delta ^k_m\delta(\vec x-\vec y),
\end{equation}
where
\begin{equation}\label{1.15}
\pi_m(x_i)=-A^j_m(\xi_i(x_k))l_j(x_i)
\end{equation}
By tedious, but direct calculation we also obtain
\begin{equation}\label{1.16}
\{\pi_m(x_i),\pi_k(y_i)\}=0,
\end{equation}
so the set $\left(\pi_m(x_i),\xi_k(y_i)\right)$ is formed by the
canonical
variables. In terms of these canonical variables the generators of
the
group of diffiomorphisms $l_j(x_i)$ has the following form:
\begin{equation}\label{1.17}
l_j(x_i)=-\frac{\partial \xi_k(x_i)}{\partial x_j}\pi_k(x_i).
\end{equation}
For the Lagrangian considered, from (\ref{1.1}) and (\ref{1.4})
follows
that $l_j(x_i)$ is given by
\begin{equation}\label{1.18}
l_j(x_i)=m\rho(x_i)v_j(x_i)
\end{equation}
and the representation (\ref{1.17}) is very similar to Clebsh
parametrization  \cite{Clebsh} of the velocity. The distinction of
(\ref{1.17}) from the original Clebsh parametrization is the
appearance of
three "potentials", instead of two, for the 3-dimensional fluid. The
reason of this difference will be discussed in the end of this
section.

The canonical Hamiltonian (\ref{1.2}) should now be expressed in the
terms
of new variables $\left(\pi_m(x_i),\xi_k(y_i)\right)$. Indeed, let us
again insert  the unity
\begin{equation}\label{1.23}
1=\int d^3 x\delta(\vec x-\vec x(\xi_i))
\end{equation}
into the integrand (\ref{1.2}) and change the order of integration:
\begin{equation}\label{1.24}
H=\int d^3 x\int d^3 \xi \left(\frac{1}{2m \rho_0(\xi_i)}\vec
p^2(\xi_i,t)+f(det
A(\xi_i))\right)\delta(\vec x-\vec x(\xi_i)),
\end{equation}
Performing  the integration over $\xi$ with the help of (\ref{2}) and
(\ref{1.4}) we obtain:
\begin{equation}\label{1.25}
H=\int d^3 x \left(\frac{1}{2m\rho (x_i)}\vec l^2(x_i)+\rho
(x_i)f(\frac{1}{\rho(x_i)})\right).
\end{equation}
Making use of (\ref{1.17}) we can express $H$ via canonical variables
$\left(\pi_m(x_i),\xi_k(y_i)\right)$:
\begin{equation}\label{1.26}
H=\int d^3 x \left(\frac{1}{2m\rho (x_i)}\frac{\partial
\xi_k(x_i)}{\partial
x_j}\frac{\partial \xi_m(x_i)}{\partial
x_j}\pi_k(x_i)\pi_m(x_i)+V(\rho
(x_i))\right),
\end{equation}
where the function $V(\rho (x_i))$ for the "potential" part of the
energy was introduced in (\ref{d}).  For usual fluid or gas this term
represents  the internal energy of the fluid and it should vanish
for uniform density distribution $\rho (x_i)=\rho_{as}$. A
phenomenological expression for $V(\rho (x_i))$ could be written as
follows \cite{Zakharov}:
\begin{equation}\label{1.27}
V(\rho (x_i))=\frac{\kappa}{2\rho_0} (\delta \rho (x_i))^2 +\lambda
(\nabla \rho (x_i))^2+... \quad,
\end{equation}
where $\delta \rho (x_i)$ is the deviation of the density from its
homogeneous distribution:
\begin{equation}\label{1.28}
\delta \rho (x_i)=\rho (x_i)-\rho_{as}.
\end{equation}
The first term in (\ref{1.27}) is responsible for the  sound wave in
the
fluid ($\kappa$ is the velocity of sound), the second term in
(\ref{1.27})
describes the  dispersion of the sound waves.

In order to reveal the relation of the Hamiltonian flow, generated by
(\ref{1.26}) with geodesic flow \cite{Arnold} let us introduce the
metric
tensor $g_{jk}(x_i)$:
\begin{equation}\label{1.29}
g_{jk}(x_i)=A^m_j(\xi_i(x_k))A^m_k(\xi_i(x_k)).
\end{equation}
With this notation (\ref{1.26}) takes the form:
\begin{equation}\label{1.30}
H=\int d^3 x
\left(\frac{1}{2m}\sqrt{g(x_i)}g^{km}(x_i)\pi_k(x_i)\pi_m(x_i)+
V(\frac {1}{\sqrt{g(x_i)}}) \right).
\end{equation}
The metric tensor with upper indices $g^{jk}(x_i)$ denotes, as
usually
the inverse matrix and
\begin{equation}\label{1.31}
g(x_i)=det g_{jk}(x_i)=\frac{1}{\rho ^2(x_i)}
\end{equation}
The representation (\ref{1.30}) permit us to consider the
hydrodynamics as the geodesic flow on the dynamical manifold with
metric $g_{jk}(x_i)$ and many general properties of the hydrodynamics
could be derived from this fact (see e.g. \cite{Arnold}).

\section{Infinite-dimensional symmetry and integrals of motion.}

As we have mentioned above, the Lagrangian (\ref{17}) possesses the
invariance with respect to the "volume preserving" group of
diffeomorphisms $SDiff[D]$, where $D\subseteq R^3$. Indeed, let us
write the Lagrangian (\ref{17}) in terms of the Lagrangian density
(to simplify the equations, in this section  we shall take the
initial density $\rho_0(\xi_i)=1$ )
\begin{equation}\label{2.1}
L=\int d^3 \xi {\cal L}(\xi_i)
\end{equation}
and consider the transformations from $SDiff[D]$ of the coordinates
$\xi_i\in D$
\begin{equation}\label{2.2}
\xi_j\rightarrow \xi'_j=\phi_j(\xi_i),
\end{equation}

\begin{equation}\label{2.3}
det\frac{\partial \phi_j(\xi_i)}{\partial \xi_k}=1.
\end{equation}
Apparently, due to (\ref{2.3}) we obtain:
\begin{equation}\label{2.4}
L=\int d^3 \phi(\xi_i) {\cal L}\left(\phi(\xi_i)\right)=\int d^3 \xi
{\cal
L}\left(\phi(\xi_i)\right)
\end{equation}
and according to Noether's theorem this invariance results in the
existence of an infinite set of integrals of motion. To obtain these
integrals we first need to find the parametrization of the
transformations
(\ref{2.2}), (\ref{2.3}) in the vicinity of identity transformation:
\begin{equation}\label{2.5}
\phi_j (\xi_i)=\xi_j+\alpha_j(\xi_i).
\end{equation}
From (\ref{2.3}) follows the equation for $\alpha_j(\xi_i)$:
\begin{equation}\label{2.6}
\frac{\partial \alpha_j(\xi_i)}{\partial \xi_j}=0.
\end{equation}
Further we must explicitly take into account that the volume
preserving
diffeomorphism (\ref{2.2}) leaves the boundary of $D$ invariant. We
shall limit ourself with the case when $D$ is formed by extraction of
the
domain with the differentiable boundary given by
\begin{equation}\label{2.7}
g(\xi_i)=0
\end{equation}
from $R^3$. Physically that means that we put in the fluid the fixed
body,
the shape of which  is given by (\ref{2.7}). The condition that the
infinitesimal
diffeomorphism (\ref{2.5}) preserves $D$ in this case is
\begin{equation}\label{2.8}
g(\xi_j+\alpha_j(\xi_i))\Big|_{g(\xi_i)=0}=0,
\end{equation}
or
\begin{equation}\label{2.9}
\alpha_j(\xi_i)\nabla_j g(\xi_i)\Big|_{g(\xi_i)=0}=0.
\end{equation}
Geometrically equation (\ref{2.9}) means that the vector $\vec
\alpha(\xi_i)$ is tangent to the surface, defined by (\ref{2.7}),
because the vector $\nabla_j g(\xi_i)\Big|_{g(\xi_i)=0}$ is
proportional
to the normal $n_j(\xi_i)$ of the surface (\ref{2.7}) at the point
$\xi_i$.

From Noether's theorem we obtain that the invariance of the
Lagrangian
with respect to the transformation (\ref{2.5}) gives the following
conservation law:
\begin{equation}\label{2.10}
\frac{\partial}{\partial t} \int_{D} d^3 \xi p_m(\xi_i)\frac{\partial
x_m(\xi_i)}{\partial \xi_l} \alpha_l(\xi_i)=0,
\end{equation}
where $\alpha_l(\xi_i)$ satisfies to the conditions (\ref{2.6}) and
(\ref{2.9}). The existence of these conditions forbids to take the
variation
of the l.h.s. of (\ref{2.10}) over $\alpha_l(\xi_i)$ and obtain the
local
form
of integrals of motion. For that we need to extract from (\ref{2.6})
and
(\ref{2.9}) the integral properties of $\alpha_l(\xi_i)$.

Consider an arbitrary, single-valued,  differentiable in $D$ function
$\beta(\xi_i)$. Then the following equations are valid:
\begin{eqnarray}\label{2.11}
&0=\displaystyle\int_{D}d^3 \xi \beta(\xi_i)\frac{\partial
\alpha_j(\xi_i)}{\partial
\xi_j}=\nonumber\\
&\displaystyle=\int_{D}d^3 \xi \frac{\partial}{\partial
\xi_j} \Bigl(\beta(\xi_i)\alpha_j(\xi_i)\Bigr)-\int_{D}d^3 \xi
\alpha_j(\xi_i)\frac{\partial\beta(\xi_i)}{\partial\xi_j},
\end{eqnarray}
The first equality is valid due to condition (\ref{2.6}). Using
Stokes theorem we can transform the integral of total derivative in
the
last equality (\ref{2.11}):
\begin{equation}\label{2.12}
\displaystyle\int_{D}d^3 \xi \frac{\partial}{\partial
\xi_j} \Bigl(\beta(\xi_i)\alpha_j(\xi_i)\Bigr)=\int_{\partial D}
dS_j\Bigl(\beta(\xi_i)\alpha_j(\xi_i)\Bigr)=0,
\end{equation}
where $\partial D$ denotes the boundary of $D$. The last integral in
(\ref{2.12}) vanishes due to  condition (\ref{2.9}) because the
differential $dS_j$ is proportional to the normal vector of the
surface
$\partial D$, defined by (\ref{2.7}). From (\ref{2.11}) and
(\ref{2.12})
we conclude that
\begin{equation}\label{2.13}
\int_{D}d^3 \xi
\alpha_j(\xi_i)\frac{\partial\beta(\xi_i)}{\partial\xi_j}=0
\end{equation}
for any smooth, differentiable $\beta(\xi_i)$. Taking this property
of
$\alpha_j(\xi_i)$ into account we obtain from the conservation laws
(\ref{2.10}) that the quantities
\begin{equation}\label{2.14}
J_k(\xi_i)=p_m(\xi_i)\frac{\partial x_m(\xi_i)}{\partial \xi_k}
\end{equation}
are conserved modulo some term which is the gradient of a scalar. In
particular, that means that
\begin{equation}\label{2.15}
R_j(\xi_i)=\epsilon_{jkl}\frac{\partial}{\partial\xi_k}J_l(\xi_i)=
\epsilon_{jkl}\frac{\partial}{\partial
\xi_k}\left(p_m(\xi_i)\frac{\partial x_m(\xi_i)}{\partial
\xi_l}\right).
\end{equation}
is the integrals of motion.
Note, that as the group of invariance is infinite-dimensional, we
obtain
an infinite number of integrals of motion. With respect to Poisson
brackets
(\ref{1.3}) the $R_j(\xi_i)$'s form an algebra. This algebra could be
written in a compact form for integrated objects
\begin{equation}\label{2.16}
R[\phi]=\int d^3 \xi \phi_j(\xi_i)R_j(\xi_i),
\end{equation}
where $\phi_j (\xi_i)$ are smooth, rapidly decreasing functions. The
algebra of $R[\phi]$ induced by Poisson brackets (\ref{1.3}) has the
form:
\begin{equation}\label{2.17}
\{R[\phi],R[\psi]\}=R[curl\phi\times curl\psi]
\end{equation}
The construction of the $x$-dependent object, corresponding to
$R_j(\xi_i)$ is not an easy task, because our "averaging" with
$\delta (\vec x-\vec x(\xi_i))$ will introduce time dependence and
instead of conserved object we shall obtain a density, whose time
derivative gives a divergence of a "current". Therefore we need to
introduce another kind of averaging without explicit refereing to
the $\vec x(\xi_i)$ coordinates. For that recall that under
diffeomorphism a closed loop transforms into closed loop. Then let
us consider such a loop $\lambda$ and a surface $\sigma$ whose
boundary is $\lambda$. The integral
\begin{equation}\label{2.18}
V=\int_{\sigma} dS_j R_j(\xi_i)
\end{equation}
where the vector $dS_j$ is as usually the area element times the
vector,
perpendicular to the surface, is conserved, because of the
conservation of
$R_j(\xi_i)$. Further, from Stokes theorem we have:
\begin{equation}\label{2.19}
T=\oint_{\lambda} d \xi_j p_m(\xi_i)\frac{\partial
x_m(\xi_i)}{\partial
\xi_j}.
\end{equation}
Changing the variables in (\ref{2.19}) we obtain:
\begin{equation}\label{2.20}
T=\oint_{\Lambda}d x_j \frac{l_j(x_i)}{\rho (x_i)}=\oint_{\Lambda}d
x_j v_j(x_i),
\end{equation}
where $\Lambda$ is the image of the loop $\lambda$ under
diffeomorphism $\xi_j\rightarrow x_j(\xi_i)$. The object (\ref{2.20})
is very well known in hydrodynamics as the "circulation" and its
conservation is known as W.Thompson theorem \cite{Thompson}. The
relation of the circulation conservation with the invariance under
special diffeomorphisms  was first explicitly established in
\cite{Salmon}, though it also could be extracted from general
discussion in Appendix 2 of \cite{Arnold}. In order to avoid
confusion we should make a remark concerning the meaning of the
contour integral (\ref{2.20}). The matter is that the diffeomorphism
$\xi_j\rightarrow x_j(\xi_i)$ transforms a fixed loop $\lambda$ into
time dependent loop $\Lambda $ and while e.g. calculating the Poisson
brackets of $T$ with Hamiltonian we need to differentiate not only
the integrand, but also the contour. The explicit form of this
integral is
\begin{equation}\label{f}
T=\oint_{\Lambda}ds \frac{\partial x_m (\xi_i(s),t)}{\partial
s}v_m(x_k (\xi_i(s),t),t),
\end{equation}
where $\xi_i(s)$ is the contour $\lambda$.
In this form it clearly seen that the circulation could not be
expressed in terms only Euler variables. In next section
we shall come back to this issue and discuss the relation of the
symmetry with respect to $SDiff$ and Euler description.

Conservation of circulation is not the only consequence of
(\ref{2.10}).
Consider for example the case of 2-dimensional space. Here, instead
of the conserved vector $R_j(\xi_i)$ we shall have the conserved
scalar
\begin{equation}\label{2.21}
R(\xi_i)=\epsilon_{kl}\frac{\partial}{\partial\xi_k}J_l(\xi_i)=
\epsilon_{kl}\frac{\partial}{\partial
\xi_k}\left(p_m(\xi_i)\frac{\partial x_m(\xi_i)}{\partial
\xi_l}\right),
\end{equation}
This scalar defines the following integrals of motion:
\begin{equation}\label{2.22}
I_n=\int_D d^2 \xi R^n(\xi_i)=\int_D
d^2\xi\Biggl(\epsilon_{kl}\frac{\partial}{\partial
\xi_k}\left(p_m(\xi_i)\frac{\partial x_m(\xi_i)}{\partial
\xi_l}\right)\Biggr)^n.
\end{equation}
Changing variables in (\ref{2.22}) $\xi_j\rightarrow \xi_j(x_i)$
which is
possible, because $\xi_j(x_i)$ is a diffeomorphism of $D$, we obtain:
\begin{equation}\label{2.23}
I_n=\int_D d^2 x
\rho(x_i)\Biggl(\epsilon_{kl}A^j_k(\xi(x))A^m_l(\xi(x))
\frac{\partial p_m(\xi(x))}{\partial x_j}\Biggr)^n,
\end{equation}
where the matrix $A^j_k(\xi)$ was defined in (\ref{14}).
We can present $I_n$ as
\begin{equation}\label{2.25}
I_n=\int_D d^2 x \rho(x_i)^{1-n}\Biggl(\epsilon_{jm}
\frac{\partial}{\partial x_j}\frac{l_m(x_i)}{\rho(x_i)}\Biggr)^n.
\end{equation}
using the following
property of 2-dimensional matrix $A^j_k(\xi)$:
\begin{equation}\label{2.24}
\epsilon_{kl}A^j_k(\xi(x))A^m_l(\xi(x))=\epsilon_{jm}det A(\xi(x))=
\epsilon_{jm}\frac{1}{\rho(x_i)},
\end{equation}
Using the relation between $\vec l(x_i)$ and velocity we can rewrite
the integrals of motion in case of 2-dimensional fluid in the
following form:
\begin{equation}\label{g}
I_n=\int_D d^2 x \rho(x_i)^{1-n}(\partial_1 v_2(x_i)-\partial_2
v_1(x_i))^n.
\end{equation}
We see that the integrals $I_n$ are the functionals only of
Euler's variables. The consequences of this property will be
discussed in the next section.

In the case of 3-dimensional space we can construct the analogous
integrals of motion, integrating the products of the vector
(\ref{2.15}):
\begin{equation}\label{2.27}
K_{j_1,j_2...j_n}=\int_D d^3 \xi
R_{j_1}(\xi_i)R_{j_2}(\xi_i)...R_{j_n}(\xi_i)
\end{equation}
Changing variables  $\xi_j\rightarrow \xi_j(x_i)$ as above we shall
obtain:
\begin{equation}\label{2.28}
R_j(\xi)\rightarrow R_j(\xi(x))=
\epsilon_{jkl}A^m_k(\xi(x))A^n_l(\xi(x))\frac{\partial
p_n(\xi(x))}{\partial x_m}.
\end{equation}
In the 3-dimensional case the matrices $A^m_j(\xi(x))$ satisfy the
equation:
\begin{equation}\label{2.29}
\epsilon_{jkl}A^m_k(\xi(x))A^n_l(\xi(x))=\epsilon_{mnr}\frac{\partial
\xi_j
(x)}{\partial x_r}det A(\xi(x)),
\end{equation}
Therefore (\ref{2.28}) takes the form:
\begin{equation}\label{2.30}
R_j(\xi(x))=
\frac{1}{\rho(x_i)}\epsilon_{mnr}\frac{\partial\xi_j(x)}{\partial
x_r}\frac{\partial p_n(\xi(x))}{\partial x_m}
\end{equation}
The integrals $K_n$ in (\ref{2.27}) become
\begin{equation}\label{2.31}
K_{j_1,j_2...j_n}=\int_{D} d^3 x
\rho(x_i)R_{j_1}(\xi(x))R_{j_2}(\xi(x))...R_{j_n}(\xi(x))
\end{equation}
Apparently, due to the presence of
$\;\frac{\partial\xi_j(x)}{\partial
x_r}\;$ in (\ref{2.30}), we can not in this case  express
(\ref{2.31})
only in terms of velocity and density, i.e. the Eulerian description
does
not admit  this kind of integrals of motion.

In the 3-dimensional case there is one more integral, which does not
exist in any other dimension. Recall that the vector $J_k(\xi_i)$,
given
by (\ref{2.14}) is conserved modulo gradient, therefore the integral
\begin{equation}\label{2.32}
Q=\int_{D}d^3 \xi J_k(\xi_i)R_k(\xi_i)
\end{equation}
is conserved because $\vec R(\xi_i)=curl\vec J(\xi_i)$.
Transforming the  $\xi_j$-dependent variables to the $x_j$-dependent
ones
in (\ref{2.32}) we obtain helicity functional :
\begin{equation}\label{2.33}
Q=\int_{D}d^ x \epsilon_{jkl}p_j(\xi(x))\frac{\partial
p_k(\xi(x))}{\partial x_l}.
\end{equation}

\section{ Euler variables.}

This section we shall discuss  the relation of Lagrange and Euler
description of hydrodynamics (see also a very deep and
interesting discussion in \cite{Salmon}, there also could be found
numerous references to the earlier investigations).

In our approach,
presented in Section 3 the phase space of the 3-dimensional fluid is
6-dimensional ($\times\infty$), as it naturally follows from Lagrange
description.
This could be compared with recent papers \cite{Jackiw}, where (in
our notations) only $\rho(x_i)$ and $v_j(x_i)$ are regarded as the
coordinates in the phase space (this point of view of  could be found
also in different text books and articles) see for example
\cite{Arnold}, \cite {Zakharov}, \cite {Lanczos}.

According to the conventional point of view the state of the fluid
is determined by its velocity and density and therefore all other
variables like ours $\vec x(\xi_i)$ are not needed. Indeed,
solving the Euler equations of motion we can express the velocity
$\vec v(x_i,t)$ and the density $\rho (x_i,t)$ at time $t$ via the
initial data $\vec v(x_i,0)$ and $\rho (x_i,0)$ . Then, from the
definition of $\vec v(x_i,t)$ (\ref{3}) follows
\begin{equation}\label{1.32}
\vec v\left(x(\xi,t),t\right))=\dot {\vec x}(\xi,t).
\end{equation}
(here we have suppressed the indices of the arguments for brevity).
Apparently we can solve these equations with respect to $\vec
x(\xi,t)$,
provided we know the initial data $\vec x(\xi,0)$. Therefore it may
seem
that the variables $\vec x(\xi,t)$ are unnecessary, as it could be
obtained through the others. But the initial data $\vec x(\xi,0)$ are
half of the  canonical variables in Hamiltonian formalism. So, in our
approach we indeed need more variables. These additional variables
provide {\it the complete description} of the fluid in a sense that
solving the equations of motion we define not only the $\vec
v(x_i,t)$ and $\rho (x_i,t)$ but also the trajectories of the
particles which could not be obtained in the conventional formalism.
The situation is analogous to the rigid body rotation: here the
phase space $\Gamma$ is formed by 3 angles and 3 components of the
angular momentum $J_i$. As the Hamiltonian depends on the components
of $J_i$ only, we can consider separately the evolution of the
angular momentum. This is incomplete description for this mechanical
system. The complete one certainly should includes the evolution of
the angles, which define the location of the rigid body in space.

In the case of  fluid dynamics {\it the complete description} is to be done
in the 6-dimensional phase space $\Gamma$ formed by $\vec x(\xi)$ and $\vec
p(\xi)$ (or $\vec \xi(x)$ and $\vec l(x)$). If, as it is usually the case,
the Hamiltonian depends only on $\vec l(x)$ and $\rho(x)$, partial
description in terms of the velocities and densities considered as
"relevant" variables is possible. This means that we do not care about the
evolution of  the whole set of coordinates of the phase space which are
considered as inessential. The "relevant" part of the coordinates does not
necessarily form a simplectic subspace in $\Gamma$.  The non-degenerate
Poisson brackets in $\Gamma$ could become degenerate on the subset of
$\Gamma$, corresponding to the "relevant" variables. This is indeed the
case in the rigid body and in the conventional fluid dynamics. In the first
case the degeneracy of the algebra of Poisson brackets for the "relevant"
variables --- angular momentum is well known. Its center element is   the
Casimir operator of the rotation group.

In the case of fluid dynamics the algebra of the "relevant" variables
--- velocities and densities for arbitrary dimension has the
following form:
\begin{eqnarray}\label{1.33}
&\{v_j(x_i),v_k(y_i)\}&=-\frac{1}{m \rho(x)}\left(\nabla_jv_k(x_i)-
\nabla_kv_j(x_i)\right)\delta(\vec x-\vec y)\nonumber\\
&\{v_j(x_i),\rho(y_i)\}&=\frac{1}{m}\nabla_j\delta(\vec x-\vec
y)\nonumber\\
&\{\rho(x_i),\rho(y_i)\}&=0
\end{eqnarray}
This algebra could be obtained from equations (\ref{1.5}),
(\ref{1.10})and (\ref{1.11}). The center of this algebra is
infinite-dimensional and its structure  depends on the dimension of
$x$-space. To find the center of it the following consideration could
be useful. In the previous section we have considered the invariance
of the fluid dynamics with respect to the infinite-dimensional group
of special diffiomorphisms $SDiff$. This group acts on the
Lagrange variables, and in \cite{Salmon} it was called "relabeling
symmetry" . Apparently, this transformations do not affect the
Euler's variables $\vec v(x_i)$ and $\rho(x_i)$. Therefore these
variables should have vanishing Poisson brackets with the integrals of
motion, corresponding to the invariance with respect to the special
diffiomorphisms. Indeed, the direct calculation gives:
\begin{eqnarray}\label{h}
&\{R_{kl}(\xi_i),v_m(x_i)\}&=0, \nonumber\\
&\{R_{kl}(\xi_i),\rho(x_i)\}&=0,
\end{eqnarray}
where we denote as $R_{kl}(\xi_i)$ tensor
\begin{equation}\label{j}
R_{kl}(\xi_i)=\frac{\partial J_k(\xi_i)}{\partial
\xi_l}-\frac{\partial J_l(\xi_i)}{\partial \xi_k},
\end{equation}
which contrary to the vector $R_k(\xi_i)$ defined in the previous
section, is appropriate for any dimensions.
Needless to say that the circulations (\ref{2.20}) also commute with
$\vec v(x_i)$ and $\rho(x_i)$. This observation gives us a guideline
for the construction of the cental elements of the algebra
(\ref{1.33}). Indeed, if we could find a functional of
$R_{kl}(\xi_i)$, such that after transformation to $x$-dependent
representation it will depend only of $\vec v(x_i)$ and $\rho(x_i)$,
then it automatically falls into the center of (\ref{1.33}). As we
have seen in the previous section, in case of 2-dimensional fluid we
were lucky and the integrals $I_n$, given by (\ref{g}) were expressed
only via Euler variables. Therefore $I_n$ form the center, the fact
which is known since long , see \cite{Arnold}, \cite{Jackiw} for the
discussion. Here we want to note the remarkable fact: the symmetry,
which is unknown to the variables $\vec v(x_i)$ and $\rho(x_i)$,
helps to find the center of its algebra!

The 3-dimensional case we shall consider in details in the next
Section. The Casimirs which follow from the reasonings, given above
are the "helicity" functional
\begin{equation}\label{1.35}
Q=\int d^3 x \epsilon_{jkl}v_j(\vec x)\nabla_k v_l(\vec x).
\end{equation}
The other Casimir is the total number of particles $N$(valid for any
dimension, for $d=2,N=I_0$):
\begin{equation}\label{1.a}
N=\int d^3 x \rho(x_i)
\end{equation}

\section{ $C^2$ Hydrodynamics}

The 3-dimensional case, which is very important for applications was
considered by many authors starting from XIX century. It is hardly possible
to give an exhaustive list of references. Recently it was discussed in
\cite{Jackiw1} (see also \cite{Salmon} for earlier references) where it was
suggested to build Hamiltonian formalism for 3-dimensional Euler fluid
using Clebsh parametrization for the velocity :
\begin{equation}\label{4.1}
\vec v(x_i,t)=\vec \partial \alpha (x_i,t)+\beta \vec \partial
\gamma(x_i,t)
\end{equation}
where the new functions $\alpha(x_i,t),\beta(x_i,t),\gamma(x_i,t)$
together with density $\rho(x_i,t)$ are used for the construction of
the coordinates of the phase space. What we are going to suggest here
is an alternative approach, which has certain advantages \cite{C2}.

Let us consider a mechanical system which is described by a pair of
complex coordinates which belong to $C^2\times\infty$:\quad
$u_{\alpha}(x_i,t),\quad \bar u_{\alpha}(x_i,t)$, where $\alpha=1,2$.
The Lagrangian for this system we shall take in the following form:
\begin{eqnarray}\label{4.2}
L&=&\int d^3 x\{\frac{i m}{2}(\bar u(x_i,t) \dot u(x_i,t)-\dot{\bar
u}(x_i,t)u(x_i,t)) \nonumber \\
&+&m\frac{(\bar u(x_i,t)\vec \partial u(x_i,t)-\vec\partial\bar
u(x_i,t)u(x_i,t))^2}{8 \bar u(x_i,t)u(x_i,t)} -V(\bar
u(x_i,t)u(x_i,t))\},
\end{eqnarray}
where we assume the summation over indexes. The canonical momenta,
corresponding to the variables $u_{\alpha}(x_i),\quad \bar
u_{\alpha}(x_i)$ are given by equations (from now on we again will
suppress the time argument)
\begin{eqnarray}\label{4.3}
p^{u}_{\alpha}(x_i)&=&\frac{i m}{2}\bar u_{\alpha}(x_i), \nonumber \\
p^{\bar u}_{\alpha}(x_i)&=&-\frac{i m}{2}u_{\alpha}(x_i).
\end{eqnarray}
As it expected for the Lagrangian which is a  linear function of
velocities, the equations (\ref{4.3}) define the constraints on the
canonical variables:
\begin{eqnarray}\label{4.4}
\lambda_{\alpha}^1(x_i)=p^{u}_{\alpha}(x_i)-\frac{i m}{2}\bar
u(x_i)_{\alpha}\sim 0, \nonumber \\
\lambda_{\alpha}^2(x_i)=p^{\bar u}_{\alpha}(x_i)+\frac{i
m}{2}u(x_i)_{\alpha}\sim 0.
\end{eqnarray}
The Poisson brackets of the constraints are non-degenerate
\begin{equation}\label{4.5}
\{\lambda_{\alpha}^1(x_i),\lambda_{\beta}^2(y_i)\}=i m\delta_{\alpha
\beta}\delta(\vec x-\vec y).
\end{equation}
and we can use these constraints to eliminate canonical momenta
$p^{u}_{\alpha}(x_i),p^{\bar u}_{\alpha}(x_i)$ using Dirac procedure
\cite{Dirac}. The resulting Poisson (Dirac) brackets for the rest of
coordinates of the phase space $\tilde{\Gamma}$ are:
\begin{eqnarray}\label{4.6}
\{u_{\alpha}(x_i), \bar u_{\beta}(y_i)\}&=&\frac{i}{m}\delta_{\alpha
\beta}\delta(\vec x-\vec y),\nonumber\\
\{u_{\alpha}(x_i),  u_{\beta}(y_i)\}&=&0,\nonumber\\
\{\bar u_{\alpha}(x_i), \bar u_{\beta}(y_i)\}&=&0.
\end{eqnarray}
The Hamiltonian, corresponding to the Lagrangian (\ref{4.2}) has the
following form
\begin{equation}\label{4.7}
H=\int d^3 x[-m\frac{(\bar u(x_i)\vec \partial
u(x_i)-\vec\partial\bar u(x_i)u(x_i))^2}{8 \bar u(x_i)u(x_i)}
+V(\bar u(x_i)u(x_i))]
\end{equation}

Now we shall explain why we consider this system. Let us form the
following objects:
\begin{eqnarray}\label{4.8}
\vec v(x_i)&=&\frac{1}{2i}(\bar u(x_i)\vec \partial
u(x_i)-\vec\partial\bar u(x_i)u(x_i)),\nonumber \\
\rho(x_i)&=&\bar u(x_i)u(x_i).
\end{eqnarray}
The notations we have used here are not accidental. The point is that
if we shall calculate the Poisson brackets for (\ref{4.8}), using
(\ref{4.6}) the result will exactly coincide with (\ref{1.33}). The
Hamiltonian $H$, given by (\ref{4.7}), being expressed via $\vec
v(x_i)$ and ${\rho(x_i)}$ will take the following form:
\begin{equation}\label{4.9}
H=\int d^3 x[\frac{1}{2}m \rho(x_i)\vec v(x_i)^2+V(\rho(x_i))],
\end{equation}
which also coincides with  Hamiltonian given by (\ref{1.25}).

The equations of motion for the variable $u_{\alpha}(x_i), \bar
u_{\alpha}(x_i)$ have the usual form:
\begin{eqnarray}\label{4.10}
\dot u_{\alpha}(x_i)=\{H,u_{\alpha}(x_i)\}\nonumber\\
\dot {\bar u}_{\alpha}(x_i)=\{H,\bar u_{\alpha}(x_i)\}
\end{eqnarray}
Apparently, the correct equations of motion , including the continuity
equation for variables $\vec v(x_i)$ and ${\rho(x_i)}$ follow from
(\ref{4.10}).

As was mentioned above, the description of the fluid in terms of
$u_{\alpha}(x_i),\quad \bar u_{\alpha}(x_i)$ is rather similar to
the description which uses  Clebsh parametrization. Indeed, these
variables could be presented in the following form:
\begin{equation}\label{4.11}
u_{\alpha}(x_i)=\displaystyle\sqrt{\rho(x_i)}e^{i
\phi(x_i)/2}\left(\begin{array}{c}
e^{-i\psi(x_i)/2}cos\frac{\alpha(x_i)}{2}\\
e^{i\psi(x_i)/2}sin\frac{\alpha(x_i)}{2}
\end{array}\right)\
\end{equation}
from where we obtain the representation for the velocity through
angles $\phi(x_i), \psi(x_i)$ and $\alpha(x_i)$
\begin{equation}\label{4.12}
\vec v(x_i)=\frac{1}{2}(\vec \partial \phi(x_i)-\vec \partial
\psi(x_i)cos\alpha(x_i))
\end{equation}
These equation defines the velocity, if Clebsh parameters are known.
Also, as is well-known  (see e.g.\cite{Lamb},\cite{Jackiw1}) any
differentiable vector field $\vec v(x_i)$ has the local
representation (\ref{4.12})\quad. In other words, knowing $\vec
v(x_i)$, we can construct Clebsh parameters
$\alpha(x_i),\phi(x_i),\psi(x_i)$ with some ambiguity. This ambiguity
arises as a set of integration constants . In  our construction this
ambiguity could be understand as follows. The Lagrangian function
(\ref{4.2})  we consider is invariant with respect to the symmetry
group $U(2)$ which acts as follows:
\begin{eqnarray}\label{4.13}
u_{\alpha}(x_i)\rightarrow \tilde{u}_{\alpha}(x_i)=T_{\alpha
\beta}u_{\beta}(x_i), \quad T^{+}T=1
\end{eqnarray}
and according to Noether's theorem the integrals of motion, which is
the generators of these transformations are :
\begin{equation}\label{4.14}
t^{0}=\int d^{3}x\frac{1}{2}\bar{u}(x_i)u(x_i),\quad t^{a}=\int
d^{3}x \bar{u}(x_i)\frac{\sigma^{a}}{2} u(x_i)
\end{equation}
The transformations (\ref{4.14}) change the Clebsh variables, but does
not affect the Euler's variables. So, in particular, the constant
shift of the angle $\phi(x_i)$ is generated by ex-Casimir $N$, which
in $\tilde{\Gamma}$ has lost its status, the generator $t^{3}$ shifts
the angle $\psi(x_i)$ , the other two generators mix the angles
$\psi(x_i)$ and $\alpha(x_i)$ . So the system described by variables
$(u_{\alpha}(x_i), \bar{u}_{\alpha}(x_i))$ is a Hamiltonian system
with symmetry  and we can reduce its phase space by procedure given
by Souriau \cite{Sour} and Marsden and Weinstain \cite{MarsW} . The
reduced phase space is the space, where "live" almost all the Euler
variables. The latter means that  the  procedure of reduction implies
fixing the integrals of motion, in particular $t^{o}=\frac{N}{2}$
does not anymore belongs to the set of variables.

The only problem we have now is   the "helicity" functional, which
still the is the Casimir and the reduction does not remove it. For
finite dimensional systems the existence of Casimir implies the
degerancy of Poisson brackets. It could be easily seen from the
following consideration. By definition the Casimir $C$ should has a
vanishing brackets with all variable
\begin{equation}\label{4.15}
\{p_{k},C\}=0, \quad \{q_{k},C\}=0
\end{equation}
where $p_{k},q_{k}$ are all set of coordinates of the phase space. If
the Poisson brackets are non-degenerate, the equations (\ref{4.15})
mean that $C$ is a constant. The situation for the infinite
dimensional system is different because of existence of so called the
functionals with zero variation . Consider for example an infinite
dimensional system , which is described by the set of canonical
variables $p(x), q(x)$ where $x\in R$ . The Poisson brackets are
non-degenerate:
\begin{equation}\label{4.16}
\{p(x),q(y)\}=\delta(x-y).
\end{equation}
In this case we can easily construct a nontrivial functional, which
will have vanishing Poisson brackets with all variables $p(x), q(x)$.
It has the following form:
\begin{equation}\label{4.17}
C=\int_{-\infty}^{\infty} dx
\frac{p'(x)q(x)-p(x)q'(x)}{p^{2}(x)+q^{2}(x)}
\end{equation}
and has a meaning of a winding number for the phase of the complex
variable $a(x)=p(x)+i q(x)$, i.e. $C$ is what physicists used to call
topological charge. Note that in order for $C$ to be the Casimir it
is not necessary to impose the condition on
$a(x)|_{x\rightarrow-\infty}=a(x)|_{x\rightarrow+\infty}$ and
compactify $R$. In this case $C$ will take an integer values and
indeed will be the winding number.

The "helicity" functional has the same origin as the functional $C$
in this example. In order to see it let us introduce a unit four
vector $F_k$ \cite{FaddeevNiemi2}:
\begin{equation}
\frac{1}{\sqrt{\bar{u}(x_i)u(x_i)}}\left( \begin{array}{c}
u_1(x_i)\\u_2(x_i)\end{array}\right)=\left(
\begin{array}{c} F_1(x_i)+iF_2(x_i)\\F_3(x_i)+iF_4(x_i)\end{array}\right)
\end{equation}
This four vector maps $S^{3}\rightarrow S^{3}$, provided we impose on
the variables $\bar{u}_{\alpha}(x_i), u_{\alpha}(x_i)$ the asymptotic
conditions : $u_{\alpha}(x)\rightarrow u^{0}_{\alpha}$, when $|\vec
x|\rightarrow \infty$ and compactify $R^{3}$. The helicity functional
$Q$ given by (\ref{1.35}) could be written in the following form:
\begin{equation}
Q=\frac{1}{3}\int d^{3}x \epsilon_{abcd}\epsilon_{ijk}F_a\partial_i
F_b\partial_jF_c\partial_kF_d,
\end{equation}
which is the standard representation for the winding number of the
map $S^{3}\rightarrow S^{3}$, so called Hopf invariant. Here again we
should note that even if we neglect the asymptotic conditions on
$u_{\alpha}(x)$ together with compactification of $R^{3}$, $Q$ is
invariant with respect to local variations and therefore has
vanishing Poisson brackets with $\bar{u}_{\alpha}(x_i),
u_{\alpha}(x_i)$. So, the conclusion of these arguments is that for
infinite dimensional mechanical systems the existence of Casimirs does
not necessary implies the degeneracy of Poisson brackets, provided
these Casimirs are related to the geometric properties of the phase
space and the helicity belongs to this class of "friendly" Casimirs.

\section{Inclusion of the electromagnetic interaction. Plasma.}

We shall consider plasma as a fluid of two components, namely
electrons with mass $m$ and electric charge $(-e)$ and ions with
mass $M$ and charge $(+e)$. The coordinates of electrons we shall
denote as $\vec x(\xi_i)$, while the coordinates of ions will be
$\vec X(\Xi_i)$.  The interaction of the components of the plasma
with the electromagnetic field $A_{\mu}(x)$ is governed by the
following Lagrangian:
\begin{eqnarray}\label{3.1}
L=&L_0^{el}+L_0^{ion}+\displaystyle e\int d^3 \xi \Biggl[A_0(\vec
x(\xi_i,t))-\dot {\vec x}(\xi_i,t)\vec
A(\vec x(\xi_i,t))\Biggr]\nonumber\\
&-e\int d^3\Xi\Biggl[A_0(\vec X(\Xi_i,t))-\dot {\vec
X}(\Xi_i,t)\vec A(\vec
X(\Xi_i,t))\Biggr]\nonumber\\
&-\frac{1}{4}\displaystyle \int d^3 x F^{\mu\nu}(x,t)F_{\mu\nu}(x,t),
\end{eqnarray}
where $L_0^{el}$ and $L_0^{ion}$ are "free" Lagrangian
\begin{eqnarray}\label{3.2}
&L_0^{el}=\displaystyle\int d^3 \xi \Biggl[\frac{m\dot{\vec
x}^2(\xi_i,t)}{2}
- f^{el}(det\frac{\partial x_j(\xi_i,t)}{\partial
\xi_k})\Biggr],\nonumber\\
&L_0^{ion}=\displaystyle\int d^3 \Xi\Biggl[\frac{M\dot{\vec
X}^2(\Xi_i,t)}{2}-f^{ion}(det\frac{\partial X_j(\Xi_i,t)}{\partial
\Xi_k})\Biggr],
\end{eqnarray}
and $F_{\mu\nu}(x,t)$ denotes the electromagnetic field tensor:
\begin{equation}\label{3.3}
F_{\mu\nu}(x,t)=\partial _{\mu}A_{\nu}(x,t)-\partial
_{\nu}A_{\mu}(x,t)
\end{equation}
The advantage of Lagrangian description of fluid (plasma) is clear.
Using of the coordinates of charged particles as the fundamental variables allows to introduce  the interaction with electromagnetic field.

The Lagrangian (\ref{3.1}) possesses  $U(1)$ gauge invariance. We
shall
use the usual Hamiltonian formalism for the constraint system
\cite{Dirac}. The canonical variables are:
\begin{eqnarray}\label{3.4}
\vec x(\xi_i), &\qquad & \vec p(\xi_i)=\frac{\delta L}{\delta
\dot{\vec
x}(\xi_i)};\nonumber\\
\vec X(\Xi_i), &\qquad  &\vec P(\Xi_i)=\frac{\delta L}{\delta
\dot{\vec
X}(\Xi_i)};\nonumber\\
\vec A(x_i),&\qquad & \vec P_{em}(x_i)=\frac{\delta L}{\delta
\dot{\vec
A}(x_i)}=-\vec E(x_i);\nonumber\\
A_0(x_i),&\qquad & P_{em}^0(x_i)=0,
\end{eqnarray}
where $\vec E(x_i)$ is the electric field strength:
\begin{equation}\label{3.5}
\vec E(x_i)=-\nabla A_0(x_i)-\dot{\vec A}(x_i).
\end{equation}
The last of the equations (\ref{3.4}) is actually the primary
constraint.
The Legendre transformation of the Lagrangian (\ref{3.1}) gives us
the
canonical Hamiltonian:
\begin{eqnarray}\label{3.6}
&H=\int d^3 x \Bigl[\frac{1}{2}\left(\vec P_{em}^2 (x_i)+\vec
H^2(x_i)\right)+A_0(x_i)\nabla \vec P_{em}(x_i)\Bigr]+\nonumber\\
&\int d^3 \xi \Bigl[\frac{1}{2m}\left(\vec p(\xi_i)+e\vec
A(x(\xi_i))\right)^2-eA_0(x(\xi_i))+f^{el}(det\frac{\partial
x_j(\xi_i,t)}{\partial \xi_k})\Bigr]+ \nonumber\\
&\int d^3 \Xi \Bigl[\frac{1}{2M}\left(\vec P(\Xi_i)-e\vec
A(X(\Xi_i))\right)^2+eA_0(X(\Xi_i))+f^{ion}(det\frac{\partial
X_j(\Xi_i,t)}{\partial \Xi_k})\Bigr],
\end{eqnarray}
where $\vec H(x_i)=curl\vec A(x_i)$ is the magnetic field
strength.The
requirement of the conservation of the primary constraint
$P_{em}^0(x_i)=0$
gives the secondary constraint ( Gauss law ):
\begin{equation}\label{3.7}
\nabla_j
\left(P_{em}(x_i)\right)_j-e\left(\rho_{el}(x_i)-\rho_{ion}(x_i)
\right)=0.
\end{equation}
Now we can add to the primary constraint $P_{em}^0(x_i)=0$ the
gauge fixing condition $A_0(x_i)=0$ and eliminate these variables
from consideration. Introducing the $x$-dependent functions
instead of $\xi$ and $\Xi$-dependent, as we did in the 2-nd
section we obtain the Hamiltonian of the plasma in the following
form:
\begin{eqnarray}\label{3.8}
&H=\int d^3 x \Bigl[\frac{1}{2}\left(\vec P_{em}^2 (x_i)+\vec
H^2(x_i)\right)
+\displaystyle\frac{\left(\vec l(x_i)+e\rho_{el}(x_i)\vec
A(x_i)\right)^2}{2m\rho_{el}(x_i)}\nonumber\\
&+\displaystyle\frac{\left(\vec L(x_i)-e\rho_{ion}(x_i)\vec
A(x_i)\right)^2}{2m\rho_{ion}(x_i)}+v_{el}(\rho_{el}(x_i))+v_{ion}
(\rho_{i
on}(x_i))\Bigr],
\end{eqnarray}
where we have introduced
\begin{eqnarray}\label{3.9}
\vec l(x_i)&=\int d^3 \xi \vec p(\xi_i)\delta (\vec x-\vec
x(\xi_i))\nonumber\\
\vec L(x_i)&=\int d^3 \Xi \vec P(\Xi_i)\delta (\vec x-\vec X(\Xi_i))
\end{eqnarray}
This Hamiltonian is gauge invariant with respect to the
transformations,
generated by the constraint (\ref{3.7}). Further we impose the
Coulomb
gauge condition on the electromagnetic field
\begin{equation}\label{3.10}
\nabla_j A_j (x_i)=0
\end{equation}
and following the usual procedure will eliminate the longitudinal
components of $\vec A(x_i)$ and $\vec P_{em}(x_i)$. As a result of
the
gauge fixing, the longitudinal part of $\vec P_{em}(x_i)$ give rise
to the
Coulomb term in the Hamiltonian
\begin{equation}\label{3.11}
H_{Col}=-e^2\int d^3 x\left(\rho_{el}(x_i)-\rho_{ion}(x_i)\right)
\left(-\frac{1}{4\pi|\vec x-\vec y|}\right)
\left(\rho_{el}(y_i)-\rho_{ion}(y_i)\right).
\end{equation}
and the whole Hamiltonian takes the following form:
\begin{eqnarray}\label{3.12}
&H=H_{Col}+\int d^3 x \Bigl[\frac{1}{2}\left(\vec P_{em\bot}^2
(x_i)+\vec
H^2(x_i)\right)\nonumber\\
&+\displaystyle\frac{\left(\vec l(x_i)+e\rho_{el}(x_i)\vec
A_\bot(x_i)\right)^2}{2m\rho_{el}(x_i)}+v_{el}(\rho_{el}(x_i))
\nonumber\\
&+\displaystyle\frac{\left(\vec L(x_i)-e\rho_{ion}(x_i)\vec
A_\bot(x_i)\right)^2}{2m\rho_{ion}(x_i)}+v_{ion}(\rho_{ion}(x_i))
\Bigr],
\end{eqnarray}
where the subscript $\bot$ denotes the transverse components of
electromagnetic variables, for which the Poisson (Dirac) brackets are
given by \cite{Dirac}
\begin{equation}\label{PB}
\{P^{em}_{\bot j}(x),A_{\bot
k}(y)\}=\left(\delta_{jk}-\frac{1}{\triangle}
\partial_j\partial_k\right)\delta(\vec x-\vec y).
\end{equation}
In equilibrium
plasma the Coulomb interaction is known to be screened by the cloud
and
the residual Debye interaction is a short rang one (see e.g.
\cite{Balescu}). In \cite{FaddeevNiemi}
it was stated that Coulomb term is reduced to the local functional of
the
difference of charge densities
$\left(\rho_{el}(x_i)-\rho_{ion}(x_i)\right)$.

Following our consideration of a fluid in the Section 3, we can
introduce
the canonical coordinates $\vec{\xi}(x_i), \vec{\pi}(x_i)$ for
electron
and $\vec{\Xi}(x_i), \vec{\Pi}(x_i)$ for ion components of the
plasma:
\begin{eqnarray}\label{3.13}
\{\pi_j(x_i),\xi_k(y_i)\}=\delta_{jk}\delta(\vec x-\vec y),
\nonumber\\
\{\Pi_j(x_i),\Xi_k(y_i)\}=\delta_{jk}\delta(\vec x-\vec y),
\end{eqnarray}
for which
\begin{eqnarray}\label{3.14}
&l_j(x_i)=\displaystyle\vec{\xi}(x_i)\frac{\partial
\vec{\pi}(x_i)}{\partial x_j},\nonumber\\
&L_j(x_i)=\displaystyle\vec{\Xi}(x_i)\frac{\partial
\vec{\Pi}(x_i)}{\partial x_j}.
\end{eqnarray}
Substituting (\ref{3.14}) into (\ref{3.12}) we obtain the Hamiltonian
in
terms of the canonical variables.

The Lagrangian (\ref{3.1}) is invariant with respect to the volume
preserving diffeomorphisms of both components of plasma
separately, so we shall have in this case two sets of conservation
laws -- one for electrons, the other for ions. Applying the
procedure, which we have described in the previous section we
shall construct the conserved circulations:
\begin{eqnarray}\label{3.15}
&V^{el}_{\Lambda}=\oint_{\Lambda}d x_j \frac{l_j(x_i)}{\rho_{el}
(x_i)},\nonumber\\
&V^{ion}_{\Lambda}=\oint_{\Lambda}d x_j \frac{L_j(x_i)}{\rho_{ion}
(x_i)},
\end{eqnarray}
where $l_j(x)$ and $L_j(x)$ are given by (\ref{3.14}). In the same
way we
can construct the analogues of the integrals (\ref{2.31}) and
(\ref{2.33})
for this case. Note that in the case of plasma the equation
(\ref{1.18})
is not valid due to the presence of the electromagnetic field,
instead we
have:
\begin{eqnarray}\label{3.16}
&{\vec l}(x)=\rho_{el}(x)\left(m{\vec v}_{el}(x)-e{\vec
A}(x)\right),\nonumber\\
&{\vec L}(x)=\rho_{ion}(x)\left(M{\vec v}_{ion}(x)+e{\vec
A}(x)\right),
\end{eqnarray}
therefore the equations (\ref{3.15}) are not the circulations of the
velocities. In the same time, adding circulations $V^{el}_{\Lambda}$
and
$V^{ion}_{\Lambda}$ on the common  contour $\Lambda$ we obtain:
\begin{equation}\label{3.17}
V_{\Lambda}=V^{el}_{\Lambda}+V^{ion}_{\Lambda}=\oint_{\Lambda}d x_j
\left(mv_{el}(x)+Mv_{ion}(x)\right),
\end{equation}
so the total circulation of elections and ions
velocities is conserved even in the presence of the
electromagnetic interaction.

Summarizing we can say that the phase space of plasma $\Gamma$  in
Coulomb gauge is the space with coordinates $\vec A_{\bot}(x),
\vec P^{em}_{\bot}(x); \vec{\xi}(x_i), \vec{\pi}(x_i);
\vec{\Xi}(x_i), \vec{\Pi}(x_i)$ with Poisson brackets given by
(\ref{PB}) and (\ref{3.13}). The evolution of a state in $\Gamma$
is defined by the Hamiltonian (\ref{3.12}). In the same time we can use
incomplete description in the terms of Euler variables, but in the case
of plasma the Poisson brackets for the velocities are changed because of the
presence of electromagnetic field in equations (\ref{3.16}). For example the
Poisson brackets for the velocities of electrons take the following form:
\begin{eqnarray}\label{3.18}
&\{v_j(x_i), v_k(y_i)\}=\nonumber\\
-&\frac{1}{m \rho(x)}
\left[\partial_j v_k(x_i)-
\partial_k v_j(x_i)+\frac{e}{m}(\partial_j A_k (x)-\partial_k A_j (x))\right]\delta(\vec x-\vec y).
\end{eqnarray}
The same way are changed the Poisson brackets of the velocities of ions,
while the brackets involving densities are not changed.

\section{Gravitating gas}

In the recent paper  \cite{Pronko} we have considered the Hamiltonian
formalism for fluid and gas based on the Lagrangian description. It was
pointed out in this paper that apart from other advantages, the Lagrangian
description, which uses the trajectories of the particles of fluid (gas) as
the dynamical variables, is the most convenient  for the introduction of
interaction. In particular in the previous section it was demonstrated  how the
introduction of the electromagnetic interaction of particles which
constitute the fluid, provide us with the theory of plasma. In the present
paper we are going to consider in analogous way the theory of gas of
particles which interact with each other through gravitational Newton
potential. The system of such particles could be considered as a model for
the motion of stars in a galaxy when the gravitation interaction prevails
all other interaction. The total number of stars in typical galaxy is of
the order of $10^{13}-10^{14}$, so it may be reasonable to consider this
collection of "particles" as a gas.

The simplest model, which is usually used for numerical simulation of
N-body model of galaxy is described by the Hamiltonian
\begin{eqnarray}\label{5.1}
H=\sum_{i=1}^{N}\frac{\vec
p{_i}^2}{2m_i}-\gamma\sum_{i\not=j}^{N}\frac{1}{|\vec x_i-\vec x_j|} ,
\nonumber
\end{eqnarray}
where $\vec x_i, \vec p_i$ are canonical coordinates of particles (stars)
with masses $m_i$. The model we are going to consider is based on the
dynamics described by $H$ with the assumption $m_i=m$, when $N\rightarrow
\infty$. For this limit there appears a natural desire to consider  a
continuous distribution of the particles as it is done in the theory of
fluid or gas \cite{Sanchez}.

The Lagrangian of the continuous system of particles interacting through Kepler
potential has the following form:
\begin{equation}\label{5.2}
L=\int d^3 \xi \rho_0(\xi_i)\frac{m\dot{\vec x}^2
(\xi_i,t)}{2}+\frac{\gamma}{2} \int d^3 \xi d^3 \xi'
\frac{\rho_0(\xi_i)\rho_0(\xi'_i)}{|\vec x(\xi_i,t)-\vec x(\xi'_i,t)|},
\end{equation}
where $\gamma$ denotes the gravitational constant. Comparing [\ref{5.2}] with
[\ref{17}] we see that the only difference is in the form of "potential energy"
therefore the canonical formalism for gravitating gas essentially the same ,
as for fluid. In such a way we can immediately write the Hamiltonian of gravitating
gas:
\begin{equation}\label{5.3}
H=\int d^3 \xi\frac{\vec p(\xi,t)^2}{2m}-\frac{\gamma}{2}\int d^3\xi
d^3\xi'\frac{\rho_0(\xi)\rho_0(\xi')}{|\vec x(\xi_i,t)-\vec x(\xi'_i,t)|}
\end{equation}
Inserting into the integrals in the r.h.s of [\ref{5.3}] the unity
\begin{equation}
1=\int d^3 x\delta(\vec x-\vec x(\xi_i)),
\end{equation}
as we did in the {\bf Section 3} we can rewrite [\ref{5.3}] via $x$-dependent
variables:
\begin{equation}\label{5.4}
H=\int d^3 x \frac{1}{2m\rho (x_i)}\vec l^2(x_i)-\frac{\gamma}{2}\int d^3 xd^3
y\frac{\rho(x_i)\rho(x_i)}{|\vec x-\vec y|}.
\end{equation}
Certainly the attractive interaction between particles will
dramatically change the properties of gravitating gas

The equations of motion,
which follow from the Lagrangian (\ref{5.2}) have the form:
\begin{equation}\label{15}
m\ddot{\vec x} (\xi_i,t)+\gamma\int d^3 \xi'\rho_{0}(\xi'_i) \frac{\vec
x(\xi_i,t)-\vec x(\xi'_i,t)}{|\vec x(\xi_i,t)-\vec x(\xi'_i,t)|^3}=0
\end{equation}
Translating equation (\ref{15}) on to the language of Euler variables, as
has been done above we arrive at the following set, including the
continuity equation:
\begin{eqnarray}\label{16,17}
\dot{\vec v}(x_i,t)+ v_k(x_i,t)\displaystyle\frac{\partial}{\partial x_k}
\vec v (x_i,t) &=&\frac{\gamma}{m}\displaystyle\frac{\partial}{\partial\vec
x}
\int d^3 y \frac{\rho(y_i,t)}{|\vec x-\vec y|}, \\
\dot \rho (x_i,t)+\displaystyle\frac{\partial}{\partial \vec x}\Bigl(\rho
(x_i,t)\vec v (x_i,t)\Bigr)&=&0.
\end{eqnarray}
Note that the r.h.s of the equation (16) in the case of ordinary gas or
fluid is expressed through the internal pressure $p(x_i)$;
\begin{equation}\label{18}
\dot{\vec v}(x_i,t)+ v_k(x_i,t)\displaystyle\frac{\partial}{\partial x_k}
\vec v (x_i,t)=-\frac{1}{\rho (x_i)}\displaystyle\frac{\partial}{\partial
\vec x}p(x_i)
\end{equation}
The set (16),(17) defines the evolution of initial distribution of
$\rho(x_i,t_0), \vec v(x_i,t_0)$ of gravitating gas and, besides it could
be used to find the static configuration of this gas for different boundary
conditions. In particular we can explore the possibility of the existence
of the static isolated configurations of gravitating gas. Isolation here
means that density $\rho(x_i)$ vanishes at infinity. Note, that  for usual
fluid, gas or plasma these kind of solutions are forbidden due to virial
arguments, known in the case of plasma as Shafranov's theorem
\cite{Shaf},\cite{Fad1}. For the gas, describing by equation (\ref{18})
this theorem could be proven as follows. First, using the continuity
equation let us rewrite equation (\ref{18}) for static case in the
following form:
\begin{equation}\label{19}
\displaystyle\frac{\partial}{\partial
x_k}\biggl[\rho(x_i)v_k(x_i)v_j(x_j)+\delta_{jk}p(x_i)\biggr]=0.
\end{equation}
Integrating (\ref{19}) with $x_j$ over $R^3$ we obtain:
\begin{eqnarray}\label{20}
&0=\int d^3 x x_j \displaystyle\frac{\partial}{\partial
x_k}\biggl[\rho(x_i)v_k(x_i)v_j(x_j)+\delta_{jk}p(x_i)\biggr]=\\
\nonumber &\int d^3 x \displaystyle\frac{\partial}{\partial
x_k}\biggl(x_j\biggl[\rho(x_i)v_k(x_i)v_j(x_j)+\delta_{jk}p(x_i \biggr]
\biggr)-\\ \nonumber &\int d^3 x
\delta_{jk}\biggl[\rho(x_i)v_k(x_i)v_j(x_j)+\delta_{jk}p(x_i)\biggr]
\end{eqnarray}
For usual gases $p(x_i)\sim \rho^{\gamma}(x_i), \gamma >0$, therefore the
integral over divergence will vanish for isolated solutions for which
$\rho(x_i)\rightarrow 0$ when $|\vec x|\rightarrow \infty$ and we the
obtain the following equation:
\begin{equation}\label{21}
\int d^3 x \biggl[\rho(x_i)\vec v^2(x_j)+3 p(x_i)\biggr]=0.
\end{equation}
Apparently, this equation could be satisfied only for the case
$\rho(x_i)=0$, i.e. there is no isolated in the above formulated sense,
static solutions of the equation (\ref{18}). In order to have a static
solution of (\ref{18}) we need to change the boundary condition
$\rho(x_i)\rightarrow 0$ to the condition $\rho(x_i)\rightarrow \rho_{as}$
, where $\rho_{as}$ is asymptotic uniform density.

Now we shall show that the arguments of this theorem bring no obstacles for
gravitating gas. For this  we again will rewrite the equations (16) for the
static case, using continuity equation in the following form:
\begin{equation}\label{22}
\biggl[\frac{\partial}{\partial x_j}\biggl(\rho(x_i)v_k(x_i)v_j(x_i)\biggr)
-\frac{\gamma}{m}\rho(x_i)\frac{\partial}{\partial x_k}\int d^3 y
\frac{\rho(y_i,t)}{|\vec x-\vec y|}\biggr]=0,
\end{equation}
Integrating (\ref{22}) with $x_k$ over $R^3$ we obtain:
\begin{equation}\label{23}
\int d^3 x x_k \biggl[\frac{\partial}{\partial
x_j}\biggl(\rho(x_i)v_k(x_i)v_j(x_i)\biggr)
-\frac{\gamma}{m}\rho(x_i)\frac{\partial}{\partial x_k}\int d^3 y
\frac{\rho(y_i,t)}{|\vec x-\vec y|}\biggr]=0.
\end{equation}
Consider the first term of the integrand in (\ref{23}). Integrating by
parts  as above and taking into account the asymptotic conditions for
$\rho(x_i)$ we obtain:
\begin{equation}\label{24}
\int d^3 x x_k \displaystyle\frac{\partial}{\partial
x_j}\biggl(\rho(x_i)v_k(x_i)v_j(x_i)\biggr)=-\int d^3x \rho(x_i)\vec
v^2(x_i).
\end{equation}
Integration of the second term of the integrand in (\ref{23}) is
straightforward, yielding
\begin{equation}\label{25}
-\frac{\gamma}{m}\int d^3 x x_k \rho(x_i)\frac{\partial}{\partial x_k}\int
d^3 y \frac{\rho(y_i,t)}{|\vec x-\vec y|}=\frac{\gamma}{2m}\int d^3 xd^3
y\frac{\rho(x_i)\rho(x_i)}{|\vec x-\vec y|}
\end{equation}
In such a way from equation (\ref{23}) we obtain:
\begin{equation}\label{26}
\int d^3x \rho(x_i)\vec v^2(x_i)-\frac{\gamma}{2m}\int d^3 xd^3
y\frac{\rho(x_i)\rho(x_i)}{|\vec x-\vec y|}=0.
\end{equation}
This relation apparently could be satisfied for non-trivial configurations
of $\rho(x_i),\vec v(x_i)$. The energy functional, corresponding to the
Lagrangian (\ref{14}) has the following form:
\begin{equation}\label{27}
E=\frac{m}{2}\int d^3x \rho(x_i,t)\vec v^2(x_i,t)-\frac{\gamma}{2}\int d^3
xd^3 y\frac{\rho(x_i,t)\rho(y_i,t)}{|\vec x-\vec y|}
\end{equation}
The two terms of the energy functional have clear interpretation as kinetic
$T$ and potential $U$ parts of energy and equation (\ref{23}) expresses
famous "virial theorem" \cite{Landau}:
\begin{equation}\label{28}
2T=-U.
\end{equation}
Note here, that in the "virial theorem" equation (\ref{28}) holds true for
mean values of kinetic and potential energies, while in our case of static
solutions there is no need to average over time.

Using the relation (\ref{28}) we can easily find the total energy of static
configuration of the gravitating gas:
\begin{eqnarray}\label{29}
E_{static}=&-\displaystyle\frac{m}{2}\int d^3x \rho(x_i)\vec
v^2(x_i)=\\
\nonumber =&-\displaystyle\frac{\gamma}{4}\int d^3 xd^3
y\frac{\rho(x_i,t)\rho(y_i,t)}{|\vec x-\vec y|}.
\end{eqnarray}
So, the total energy of the static solution is negative, as for the bound
state of Kepler problem.

\section{Properties of Static Solutions}

The equations which define our static configurations of gravitating gas
have the following form:
\begin{eqnarray}\label{30}
v_k(x_i)\displaystyle\frac{\partial}{\partial x_k} \vec v (x_i)
&=&\frac{\gamma}{m}\displaystyle\frac{\partial}{\partial\vec x}
\int d^3 y \frac{\rho(y_i)}{|\vec x-\vec y|}, \nonumber\\
\displaystyle\frac{\partial}{\partial \vec x}\Bigl(\rho (x_i)\vec v
(x_i)\Bigr)&=&0
\end {eqnarray}
Taking divergence of the first equation :
\begin{eqnarray}\label{31}
\vec\partial \Bigl(v_k(x_i)\displaystyle\frac{\partial}{\partial x_k}\vec v
(x_i)\Bigr) &=&\frac{\gamma}{m}\Delta
\int d^3 y \frac{\rho(y_i)}{|\vec x-\vec y|}= \nonumber\\
&=&\frac{\gamma}{m}(-4\pi)\rho(x_i),
\end{eqnarray}
 we can write the whole set of the equations for static
configuration in a pure local form:
\begin{eqnarray}\label{32}
\vec\partial \Bigl(v_k(x_i)\displaystyle\frac{\partial}{\partial
x_k}\vec v (x_i)\Bigr)&=&-4\pi\frac{\gamma}{m}\rho(x_i),\nonumber\\
\displaystyle\frac{\partial}{\partial \vec x}\Bigl(\rho
(x_i)\vec v (x_i)\Bigr)&=&0,\nonumber\\
\vec\partial \times \Bigl(v_k(x_i)\displaystyle\frac{\partial}{\partial
x_k}\vec v (x_i)\Bigr)&=&0,
\end{eqnarray}
where the last equation  requires the expression
$v_k(x_i)\frac{\partial}{\partial x_k} \vec v (x_i)$ to be a gradient.

Now we are going to derive an important inequality, which bounds the
$(-E_{static})$ of any solution of (\ref{32}) from below. For this let us
introduce the notation for the potential $U(x_i)$:
\begin{equation}\label{33}
U(x_i)=-\int d^3 y \frac{\rho(y_i)}{|\vec x-\vec y|}
\end{equation}
Integrating by parts we obtain the following relation:
\begin{equation}\label{34}
\int d^3 x \Bigl(\vec\partial U(x_i)\Bigr)^2=4 \pi\int d^3 xd^3
y\frac{\rho(x_i,t)\rho(y_i,t)}{|\vec x-\vec y|}
\end{equation}
Using (\ref{34}) we can write the expression for the energy  given by
 (\ref{29}) of any static configuration, which is the solution of
(\ref{32}) in the following form:
\begin{equation}\label{35}
E_{static}=-\frac{\gamma}{16 \pi}\int d^3 x \Bigl(\vec\partial
U(x_i)\Bigr)^2
\end{equation}
Substituting into (\ref{35}) the expression for the gradient of the
potential $U(x_i)$ from the first equation (\ref{30}), we obtain the
expression for the $E_{static}$ only through the field of velocity:
\begin{equation}\label{36}
-E_{static}=\frac{m^2}{16 \pi \gamma}\int d^3 x
\Bigl(v_k(x_i)\displaystyle\frac{\partial}{\partial x_k}\vec v
(x_i)\Bigr)^2
\end{equation}
This form is most convenient for the derivation of desired inequality. Now
let us consider a function $f(x_i)$ from the Hilbert space $W_2 ^1$, which
consists of all measurable functions on $R^3$, which have at least one
derivative and square integrable on $R^3$ together with its derivatives. In
particular we assume that the density $\rho(x_i)$ belongs to $W_2 ^1$.
Taking into account the equation (\ref{31}) we have
\begin{eqnarray}\label{37}
&|\int d^3 x \rho(x_i)f(x_i)|=|\frac{m}{4 \pi \gamma}\int d^3 x
f(x_i)\partial_k \Bigl(v_j(x_i)\partial_j v_k(x_i)\Bigr)|\nonumber \\
&=|\frac{m}{4 \pi \gamma}\int d^3 x \partial_k f(x_i)
v_j(x_i)\partial_j v_k(x_i)|\nonumber \\
&\leq\frac{m}{4 \pi \gamma}\Bigl(\int d^3 x (\partial_k f(x_i))^2
\Bigr)^{1/2}\Bigl(\int d^3 x (v_j(x_i)\partial_j v_k(x_i))^2 \Bigr)^{1/2},
\end{eqnarray}
where on the last step we have used Cauchy inequality. From (\ref{37}) we
immediately obtain the inequality:
\begin{equation}\label{38}
-E_{static}\geq \pi \gamma \frac{\Bigl(\int d^3 x
\rho(x_i)f(x_i)\Bigr)^2}{\int d^3 x(\partial_k f(x_i))^2},
\end{equation}
which is valid for any $f(x_i)$ from $ W_2 ^1$ and is saturated for
$f(x_i)=U(x_i)$. Indeed, let us calculate the derivative of the functional
in the r.h.s of (\ref{38}) with respect to $f(x_i)$:
\begin{eqnarray}\label{39}
&\displaystyle\frac{\delta}{\delta f(x_i)}\frac{\Bigl(\int d^3 x
\rho(x_i)f(x_i)\Bigr)^2}{\int d^3 x(\partial_k
f(x_i))^2}\nonumber\\
&\displaystyle=2\frac{\Bigl(\int d^3 x \rho(x_i)f(x_i)\Bigr)^2}{\int d^3
x(\partial_k f(x_i))^2}\Bigl[\frac{\rho(x_i)}{(\int d^3 x
\rho(x_i)f(x_i)}+\frac{\Delta f(x_i)}{\int d^3 x(\partial_k
f(x_i))^2}\Bigr].
\end{eqnarray}
This derivative vanishes for $f(x_i)=f_{max}(x_i)$ given by
\begin{equation}\label{40}
f_{max}(x_i)=C \Delta^{-1}\rho(x_i)=C' U(x_i),
\end{equation}
where $C$ and $C'$ are inessential constants, which do not enter into the
functional. It is easy to prove that the second variation of this
functional is negative on the $f_{max}(x_i)$, so this function provides the
absolute maximum for the functional and due to the equations
(\ref{34}-\ref{36}) its value coincides with the l.h.s. of (\ref{38}).
However, the function $f_{max}(x_i)$ does not belong to the Hilbert space
$W_2 ^1$, because the potential $U(x_i)$ has the asymptotic behavior
$\frac{1}{|\vec x|}$ at infinity and therefore is not square integrable.

The integral $J$ which enters into inequality (\ref{38}) could be written
in the following form
\begin{equation}\label{41}
J=\int d^3 x \rho(x_i)f_{N}(x_i),
\end{equation}
where we denoted as $f_{N}(x_i)$ the normalized function $f(x_i)$:
\begin{equation}\label{42}
f_{N}(x_i)=\displaystyle \frac{f(x_i)}{\sqrt{{\int d^3 x (\partial_k
f(x_i))^2}}}=\frac{f(x_i)}{\|\partial_k f(x_i)\|_2}.
\end{equation}
In the Hilbert space $W_2 ^1$ the H\"older's inequality holds true:
\begin{equation}\label{43}
|\int d^3 x f(x_i)g(x_i)|\leq
\|f\|_{p}\|g\|_{p'},\qquad\frac{1}{p}+\frac{1}{p'}=1
\end{equation}
as well, as the remarkable Ladyjenskaya's \cite{Lad} inequality:
\begin{equation}\label{44}
\|f\|_{6}\leq (48)^{1/6}\|\partial_k f\|_{2}
\end{equation}
Here we use the standard notations:
\begin{eqnarray}
\|f\|_{p}&=&\Bigl(\int d^3 x f(x_i)^{p}\Bigr)^{1/p},\nonumber\\
\|\partial_k f\|_{p}&=&\Bigl(\int d^3 x |\partial_k
f(x_i)|^{p}\Bigr)^{1/p}\nonumber
\end{eqnarray}
From these two inequalities we obtain the following bound for the integral
$J$:
\begin{equation}\label{45}
|J|\leq (48)^{1/6}\|\rho\|_{6/5}
\end{equation}
The facts mentioned above could possibly support the statement that the
($-E_{static}$) is bounded from below by appropriate norm of the density.
Indeed, the absolute maximum of the functional $|J|$ should be bigger when
its maximum in a restricted space like $W_2 ^1$, given by (\ref{45}).
However we can not present the rigorous proof of this statement.

One of the other general property of a static configurations of gas or
fluid is the existence of the topological charge --- "helicity" (or Hopf
invariant), which explicit form is
\begin{equation}\label{46}
q=\int d^3 x \vec v(x_i) rot\vec v(x_i).
\end{equation}
This object is not only the integral of motion of the equations (16) but it
also is the central element of the algebra of Poisson brackets of $(\vec
v(x_i),\rho(x_i))$ \cite{Pronko}. The existence of such an object brings
additional argument for the stability of the solitons. The role of
"helicity" in the case of the solitons in plasma was pointed out in
\cite{FaddeevNiemi2},\cite{Fad1}. Moreover, in \cite{Vac} it was shown that there
exists a remarkable inequality which bounds the energy of plasma solitons
from below by $q^{3/4}$. The derivation of such inequality for our case
(and in general for fluid solitons ) is highly desirable and we going to
consider this question in the future publications.

\section{On the Possible Structure of the Static Solutions}

The equations (\ref{32}) which define the static configurations of
gravitating gas are 3-dimensional nonlinear partial differential equations
and the probability to find an analytic solution is very low. The only case
where a class of solutions was found in a similar situation is the
t'Hooft-Polaykov monopole, but there the requirement of the spherical
symmetry simplified essentially the problem. In our case we can not expect
the spherically symmetric solution because the continuity equation requires
the trajectories of the particles be closed in order to provide the static
configuration for $\vec v(x_i), \rho(x_i)$. The simplest and most symmetric
configuration we could expect for our case is the toroidal structure, where
the density is concentrated in the vicinity of the axis of the toroid,
while the field of velocity is tangential to the embedded one into the
other toroidal surfaces. It is not the first time the toroidal-shape
soliton appeared in the context of the theory of continuous media. Since
the pioneer works of Lord Kelvin in XIX century to the present time it was
studied  by many scientists both mathematician and physicists and recently
the interest to the subject was again attracted by the works of Faddeev and
Niemi \cite{FaddeevNiemi},\cite{FaddeevNiemi2}

It is clear that in the case of attractive gravitational interaction the
particles, which constitute the gas move around the region with bigger
density (like planets around the Sun) so there should exists a collective
motion in the approximation when the radius of torus tends to infinity and
we can speak about cylindrical rather when toroidal configuration. Indeed,
let us consider this axially symmetric tornado-like solution of (\ref{32}).
For that we shall write the Ansazt:
\begin{eqnarray}\label{47}
\vec v(x_i)&=&v(r)(-\frac{y}{r},\frac{x}{r},0),\nonumber\\
\rho(x_i)&=&\rho(r),
\end{eqnarray}
where $r=\sqrt{x^2+y^2}$. The second and the third equations (\ref{32}) are
satisfied by (\ref{47}), while the first gives
\begin{equation}\label{48}
\frac{1}{r}\partial_r v(r)^2=4\pi\frac{\gamma}{m}\rho(r).
\end {equation}
This solution shows that the density stays an arbitrary function (what is
expected for the partial differential equations) and the velocity grows
with radius up to its asymptotic value. The later is the consequence of the
approximation--- here we actually have 2-dimensional potential $log r$
instead of Coulomb $\frac{1}{r}$. This example is to demonstrate that the
particles which constitute the gravitating gas in their collective motion
form the localized object.

The traditional way to tackle 3-dimensional gas or fluid is to introduce
for the velocity Clebsh parametrization suggested in \cite{Clebsh} and
recently discussed in\cite{Jackiw}. In general case this parametrization
has the following form:
\begin{equation}\label{49}
\vec v(x_i)=f(x_i)\vec \partial g(x_i)+\vec \partial h(x_i),
\end{equation}
where $f(x_i),g(x_i),h(x_i)$ are scalar functions. For the toroidal
solution with $z$ -- as the axis of symmetry we shall assume that the
density $\rho (x_i)$ does not depend upon the azimuth angle $\phi$ and
Clebsh parametrization takes the form:
\begin{equation}\label{50}
\vec v(x_i)=cos\alpha (r,z)\vec \partial \beta (r,z)+k\vec \partial \phi
(x_i),
\end{equation}
where $r,z,\phi$ are cylindrical coordinates,
\begin{eqnarray}
r&=&\sqrt{x^2+y^2}\nonumber\\
\phi(x_i)&=&arctang\frac{y}{x}\nonumber
\end{eqnarray}
In fluid dynamics the parametrization (\ref{50}) has an interesting
mechanical interpretation which we shall discuss elsewhere. The "helicity"
functional for this case has the following form
\begin{eqnarray}\label{51}
q=&\int k d \phi(x_i)\wedge
d cos\alpha(r,z)\wedge d\beta(r,z)\nonumber\\
=&2\pi k \int d cos\alpha(r,z)\wedge d\beta(r,z)
\end{eqnarray}
The function $\beta(r,z)$ in (\ref{50}) should has a singularity at the
point $(r_c,0)$ in the semi-plane $(r,z)$  where $r_c$ is the radius of axis
line of the torus. When the point $(r,z)$ goes around $(r_c,0)$, the
function $\beta(r,z)$ increases on $2\pi$\footnote{Compare this with the
similar parametrization of the dynamical variables in \cite{FaddeevNiemi2} for the
case of plasma.}. The gradient of such function will be the vector
tangential to the toroidal surfaces. The addition of  the gradient of
azimuthal angle $\phi$ makes the velocity (\ref{50}) winding also in the
azimuth direction. Such a behavior of the field of velocity leads to the
nontrivial Hopf invariant, which characterizes the homotopy classes
$\pi_3(S^2)$. Described this way the Clebsh parameters, together with the
density $\rho(x_i)$ could be found by numerical integration of the
equations (\ref{32}).

The construction we presented and discussed above would be incomplete and
too academic without an example of the galaxy which indeed may demonstrate
the toroidal structure. Fortunately such galaxy does exists. It was
discovered back in 1950 by Art Hoag and recently a very good picture was
obtained by Hubble telescope (see figure 1). On this picture clearly seen the toroidal structure.  In the gallery of the galaxies which is available on
the sites http://www.astronomy.com or http://hubblesite.ogr we can find
some other examples with the form more or less close to the torus, but the
Hoag's object is the best of all.

\begin{figure}
\centering
\includegraphics*[width=10cm,bb=0 0 701 668]{Hoag'sObject.eps}
\caption{The above photo taken by the Hubble Space
Telescope in July 2001 reveals unprecedented details of Hoag's Object
and may yield a better understanding. Hoag's Object spans about
120,000 light years and lies about 600 million light years away
toward the constellation of Serpens.}
\end{figure}

\section{Concluding remarks}

Here we have considered different aspects of Hamiltonian formalism for several types of continuous media -- gas, fluid, plasma and gravitational gas. In spite of different properties of these media, all of it could be treated within universal formalism and difference arises due to specific interaction between constituents. As it was shown the Lagrangian approach provides the simplest way to introduce different types if interaction because it uses coordinates of particles as basic variables. Having constructed Lagrangian formalism we can derive Euler equations of motion and also the whole set of Poisson brackets for Euler variables.

The $C^2$ formalism of continuous media considered in Section 6 gives a frameworks for quantum theory which may be useful for the description of Bose-Einstein condensation in gas and this subject needs further in-depth study.

Another subject which we are going to explore in the future is the theory of gravitating gas. Real challenge here is to find analytically toroidal soliton which could describe ring galaxies.

\section*{Acknowledgments}
The author is grateful to professors A.K.Likhoded and A.V.Razumov
for their comments and fruitful discussions. This work  was supported by  RFFI grant 07-01-00234.


\vspace*{0.4cm}
\begin{thebibliography}{9}
\bibitem{Landau}
L.Landau and E.Lifshitz, Fluid Mechanics , Pergamon, Oxford UK, 1987
\bibitem{Chen}
F.Chen, Introduction to Plasma Physics., 2nd edition, Plenum Press,
1983
\bibitem{Lamb}
H. Lamb, Hydrodynamics, New York, Dover, 1932
\bibitem{Arnold}
V.I.Arnold, Mathematical Methods in Classical Mechanics, New York,
Springer-Verlag 1978; V.I.Arnold, B.Khesin, Topological Methods in
Hydrodynamics, Springer-Verlag, Berlin,1998
\bibitem{Mars}
Marsden. J., and Weinstein, A.,  {\em Physica} {\bf 4D} , p. 394
,1982;
D. D. Holm, B. Kupershmidt and C.D. Levermore, {\it Phys. Lett. A}
{\bf
98} p. 389, 1983;
Marsden, J., and Weinstein, A., {\em Physica} {\bf 7D} (1983), p.305,
1983;
Holm. D.D., Marsden, J.E., Ratiu, T., and Weinstein, A.,  {\em Phys.
Reports} {\bf
123}, nos. 1 \& 2, (1985),p. 1, 1985
\bibitem{Pronko}
I. Antoniou, G.Pronko, Theor. Math. Phys. , v.141(3),1670-1685, 2004;
\bibitem{DubrovinNovikov}
B.A.Dubrovin, S.P.Novikov, Uspekhi Matematicheskikh Nauk, v.44, N4,
p.29,
1989 \{in russian\};
B.A.Dubrovin Funkcionalnii Analis i ego Prilogeniya, v.23, N2, p.57,
1989;
\{in russian\}
\bibitem{Clebsh}
A.Clebsh, J.Reine Angew.Math., v.56, p.1, 1859
\bibitem{Zakharov}
V.E.Zakharov Sov.Phys.JETP, v.60, N 5, p.1714, 1971 \{in russian\}
\bibitem{Thompson}
W.Thompson, Mathematical and Physical Papers, vol. 4, Cambridge
University
Press, Cambridge 1910;
\bibitem{Jackiw}
R.Jackiw, "Lectures on fluid mechanics" , e-preprint physics/0010042,
2000
\bibitem{Jackiw1}
R. Jackiw, V. P. Nair, S.-Y. Pi, A. P. Polychronatos, J.Phys.A, v.
37, R327-R432, 2004
\bibitem{Lanczos}
C.Lanczos, The Variational Principles of Mechanics, University of
Toronto,
Toronto, 1970
\bibitem{Salmon}
R.Salmon, Ann.Rev.Fluid Mech. v. 20, p.225, 1998
\bibitem{C2}
G.Pronko, Theor. Math. Phys., v. 148 (1), 980-985, 2006
\bibitem{Sour}
J.-M. Souriau, Maitrises de Mathematiques. Paris: Dunod. XXXII, p.
414, 1970
\bibitem{MarsW}
J. Marsden, A. Weinstain, Rep. Math. Phys. , v. 5, N 1, p. 121-130,
1972
\bibitem{Dirac}
P.A.M.Dirac. Lecture on Quantum Mechanics, Academic Press, N.Y.,
1964.
\bibitem{Balescu}
R.Balescu Equivibrium and Nonequilibrium Statistical Machanics,
Willey-Interscience Pub. 1975
\bibitem{FaddeevNiemi}
L.D.Faddeev, A.Niemi, Magnetic Geometry and the Confinement of
Electrically Conducting Plasmas, e-preprint physics/0003083, 2000
\bibitem{FaddeevNiemi2}
L.D.Faddeev, A. J. Niemi , Toroidal Configurations as stable
solitons, e-preprint hep-th/9705176,(1997)
\bibitem{Sanchez}
B. Semelin, N. Sanchez, H.J.De Vega, Self-gravitationg fluid
dynamics,instabilities and solitons. Phys. Rev. , Vol. D 63., 084005,
(2001)
\bibitem{Shaf}
V.D.Shafranov. in Reviews of Plasma Physics, vol. II, M.A. Leontovich
(editor), Consultants Bureau, New York, (1966)
\bibitem{Fad1}
L.D. Faddeev, L. Freyhult, A.J. Niemi, P.Rajan, Shafranov's virial
theorem and magnetic plasma confinement, J.Phys. A35 (2002)
L133-L140
\bibitem{Lad}
O.A. Ladyjenskaya, Mathematical problems of the dynamics of viscous
incompressible fluid, Moscow, Nauka, 1970, (in russian).
\bibitem{vac}
A.F.Vakulenko, L.V.Kapitanskii, Sov.Phys.Dokl. A495,p.499, 1979,
\end{thebibliography}
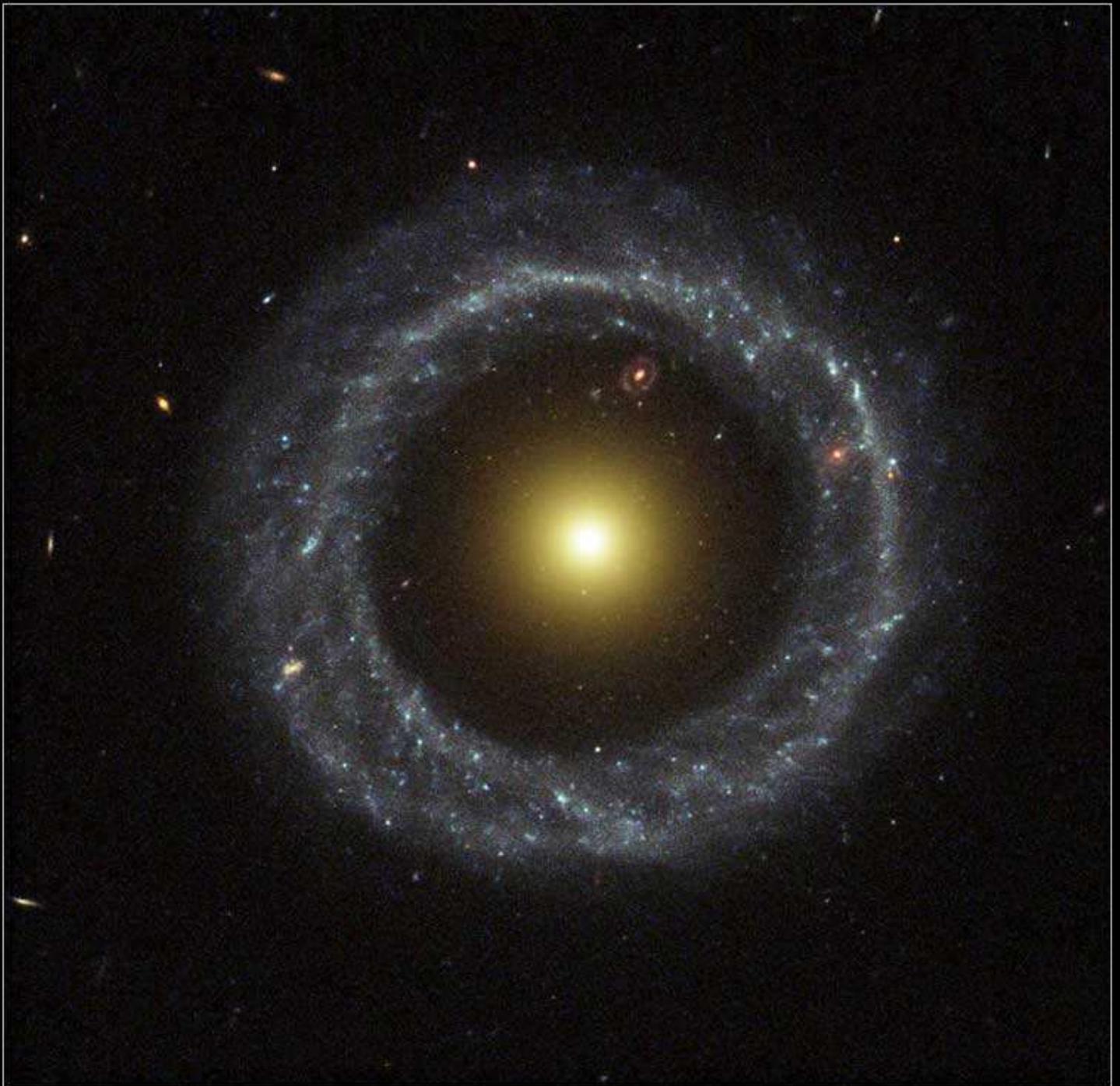
\end{document}